\documentclass[12pt,preprint]{aastex}
\usepackage{natbib}
\usepackage{xcolor}
\newcommand{\eps}[1]{\mbox{log~$\epsilon$(#1)}} 
\newcommand\species[2]{#1 {\sc #2}}
\newcommand\iso[2]{$^{\rm #1}$#2}

\def\ie{\mbox{i.e.}}
\def\eg{\mbox{e.g.}}

\def\teff{\mbox{T$_{\rm eff}$}}
\def\logg{\mbox{log~{\it g}}}
\def\vmicro{\mbox{$\xi_{\rm t}$}}
\def\kmsec{\mbox{km~s$^{\rm -1}$}}
\def\hdeight{\mbox{HD 84937}}

\shorttitle{Fe-Group Abundances}
\shortauthors{Sneden et al.}

\begin{document}

\title{Iron-Group Abundances in the Metal-Poor Main Sequence Turnoff
       Star HD~84937}

\author{
Christopher Sneden\altaffilmark{1},
John J. Cowan\altaffilmark{2}, 
Chiaki Kobayashi\altaffilmark{3},
Marco Pignatari\altaffilmark{4,5},
James E. Lawler\altaffilmark{6},
Elizabeth A. Den Hartog\altaffilmark{6},
Michael P. Wood\altaffilmark{6,7}
}

\altaffiltext{1}{Department of Astronomy and McDonald Observatory, 
                 The University of Texas, Austin, TX 78712; 
                 chris@verdi.as.utexas.edu}

\altaffiltext{2}{Homer L. Dodge Department of Physics and Astronomy, 
                 University of Oklahoma, Norman, OK 73019; 
                 jjcowan1@ou.edu}

\altaffiltext{3}{School of Physics, Astronomy and Mathematics, Centre for 
                 Astrophysics Research, University of Hertfordshire, 
                 College Lane, Hatfield AL10 9AB, UK;
                 c.kobayashi@herts.ac.uk}

\altaffiltext{4}{Konkoly Observatory, Research Centre for Astronomy and 
                 Earth Sciences, Hungarian Academy of Sciences,
                 Konkoly Thege Miklos ut 15-17, H-1121 Budapest, Hungary}

\altaffiltext{5}{NuGrid collaboration, \url{http://www.nugridstars.org}}

\altaffiltext{6}{Department of Physics, University of Wisconsin-Madison,
                 1150 University Ave., Madison, WI 53706; 
                 jelawler@wisc.edu, eadenhar@wisc.edu}

\altaffiltext{7}{current address:  National Institute of Standards and 
                 Technology, Gaithersburg, MD 20899, michael.wood@nist.gov}

\begin{abstract}
We have derived new very accurate abundances of the Fe-group
elements Sc through Zn (Z~=~21$-$30) in the bright main-sequence turnoff
star \hdeight, based on high-resolution spectra covering
the visible and ultraviolet spectral regions.
New or recent laboratory transition data for 14 species of seven elements
have been used.
Abundances from more than 600 lines of non-Fe species have been combined
with about 550 Fe lines in \hdeight\ to yield abundance ratios
of high precision.
The abundances have been determined from both neutral and ionized
transitions, which generally are in agreement with each other.
We find no substantial departures from standard LTE Saha ionization balance
in this [Fe/H]~=~$-$2.32 star.
Noteworthy among the abundances are: [Co/Fe]~=~$+$0.14
and [Cu/Fe]~=~$-$0.83, in agreement with past studies
abundance trends in this and other low metallicity stars; and
$\langle$[Sc,Ti,V/Fe]$\rangle$~=~$+$0.31, which has not been noted previously.
A detailed examination of scandium, titanium, and vanadium abundances
in large-sample spectroscopic surveys reveals that they are
positively correlated in stars with [Fe/H]~$<$~$-$2; \hdeight\
lies at the high end of this correlation.
These trends constrain the synthesis mechanisms of Fe-group elements.
We also examine the GCE abundance trends of the Fe-group elements,
including a new nucleosynthesis model with jet-like explosion effects.
\end{abstract}


\section{INTRODUCTION\label{intro}}

Massive stars began producing heavy elements near the birth of our Galaxy.
The records of their nucleosynthesis contributions are written into the
chemical compositions of very metal-poor stars of the Galactic halo.
If accurate abundances can be extracted for whole element groups in these
low metallicity stars, it will be possible to better understand their 
nucleosynthetic messages.

Over the past 15 years there have been major advances in our understanding 
of the abundances of neutron-capture elements ($n$-capture, Z~$>$~30) in 
metal-poor stars.
Many of these stars have extremely detailed $n$-capture abundance
distributions, with accurate abundances for more than 30 elements.
We know that their bulk $n$-capture contents with respect to lighter elements 
vary by orders of magnitude from star to star (\eg, \citealt{sne08} and 
references therein).
Moreover, their nucleosynthetic origin in a given star can be attributed to
rapid neutron captures (the $r$-process; \citealt{thi11} and 
references therein), and/or slow neutron captures (the $s$-process;
\citealt{kae11} and references therein), and/or a variety of other processes 
that can contribute to the zoo of anomalous abundances observed in old
halo stars (\eg, \citealt{cow77}, \citealt{tra04}, \citealt{fro06b},
\citealt{wan06}, \citealt{pig08}, \citealt{far10}, \citealt{arc11}, 
\citealt{her11}).
A major reason for these insights has been the rapid increase in 
reliability of basic atomic transition data for the $n$-capture elements 
(\eg, \citealt{law09}, \citealt{sne09} and references therein).
As the transition probabilities, isotopic splittings, and hyperfine 
structures have become more reliable, the derived $n$-capture abundances 
have also become more accurate.

The same generally cannot be said for lighter elements, in particular
for the Fe-group (21~$\leq$~Z~$\leq$~30).
There have been many large-sample spectroscopic surveys of Fe-group elements
in halo stars beginning more than 30 years ago, with the spectroscopic 
quality increasing with time.
The pioneering surveys of \cite{luc85} and \cite{mcw95} have been
followed by more detailed studies by, \eg, those of \cite{cay04}, 
\cite{bar05}, \cite{yon13}, and \cite{roe14} for red giants, and that of 
\cite{coh04} for main-sequence stars. 
But the reliability of the reported abundances in any of these surveys is not
assured.
First, for many stars not all Fe-group elements are detectable with
the available spectra.
Second, even when elements can be detected, often only neutral-species 
transitions are available for analysis in typical ground-based spectra.
The ionization balances of Fe-group elements in metal-poor warmer dwarfs and 
cooler red giants usually are dominated by the ionized species;
therefore the elemental abundances are mostly based on results from
the minority species.
Third, in many instances only a handful of lines are employed to form 
individual elemental abundances.
Finally, significant questions have been raised about whether local 
thermodynamic equilibrium (LTE) can adequately describe the ionization 
equilibrium, and large upward corrections have been proposed to LTE-based 
abundances from neutral species.

Our group has been concentrating on improving the laboratory data for
Fe-group neutral and singly-ionized transitions that are of interest to 
cool-star stellar spectroscopy.
Citations to individual papers within this series will be given in 
\S\ref{newabunds}.
In each study we have used the new atomic data to re-derive abundances
in the solar photosphere and in the very metal-poor 
([Fe/H]~$\sim$~$-$2.2\footnote{
We adopt the standard spectroscopic notation \citep{wal59} that for 
elements A and B,
[A/B] $\equiv$ log$_{\rm 10}$(N$_{\rm A}$/N$_{\rm B}$)$_{\star}$ $-$
log$_{\rm 10}$(N$_{\rm A}$/N$_{\rm B}$)$_{\odot}$.
We use the definition 
\eps{A} $\equiv$ log$_{\rm 10}$(N$_{\rm A}$/N$_{\rm H}$) + 12.0, and
equate metallicity with the stellar [Fe/H] value.})
main sequence turnoff star \hdeight.
This star has been chosen because of its brightness, well-known atmospheric
parameters, and availability of good high-resolution spectra over a very
large wavelength range.  
In this paper we will finish our reconnaissance of the Fe-group elements
in \hdeight.

In \S\ref{specdata} we introduce the high-resolution spectroscopic
data sets analyzed for the solar photosphere and \hdeight.
New abundances of several elements and summaries of recent
results for other elements are given in \S\ref{newabunds}.
Interpretation and discussion of the iron-peak data and abundances,  
particularly nucleosynthesis origins and production mechanisms, 
are included in \S\ref{results}.  
The implications for Galactic chemical evolution, accompanied by 
detailed models, follow in \S\ref{gce}. 
Finally,  a summary and conclusions are detailed in \S\ref{sum}.

\section{SPECTROSCOPIC DATA AND ANALYSES\label{specdata}}

Solar and HD~84937 data sets are the same as in previous
papers of this series. 
The spectra have been discussed in detail by \cite{law13}; here we
provide a brief summary.
For the Sun we used the photospheric center-of-disk spectrum of 
\cite{del73}\footnote{
Available at http://bass2000.obspm.fr/solar\_spect.php}.
The ground-based spectrum of \hdeight\ was obtained from the 
ESO VLT UVES archive\footnote{
Under programs 073.D-0024 (PI: C. Akerman) and 266.D-5655 (Service Mode).}.
It extends over 3100~$\lesssim$~$\lambda$~$\lesssim$ 10000~\AA, with 
resolving power $R$~$\equiv$~$\lambda/\Delta\lambda$~$\sim$ 60,000, 
and signal-to-noise $S/N$~$\sim$~100 at 3500~\AA, increasing steadily at 
longer wavelengths to a maximum $S/N$~$\sim$~300.
The vacuum-UV spectrum of \hdeight\ comes from the Hubble Space Telescope 
Imaging Spectrograph (HST/STIS) public archive, having been part
of proposal \#7402 (PI: R. C. Peterson) and originally used in a study of
chromospheric emission in metal-poor main-sequence stars \citep{pet97}. 
For this observation HST/STIS used its E230M grating to produce a wavelength 
range 
2280~$\lesssim$~$\lambda$~$\lesssim$ 3120~\AA, $R$~$\sim$ 25,000, and average 
$S/N$~$\sim$~50 after co-addition of the two spectra obtained for this star.

The adopted model solar photosphere is the empirical one determined
by  \cite{hol74}, hereafter called HM.
This model was used in the first of our atomic transition papers 
\citep{law01a}, and we have stayed with this choice for all succeeding 
papers to maintain internal consistency.
Likewise, for \hdeight\ we again employ an interpolated model 
from the \cite{kur11}\footnote{ 
http://kurucz.harvard.edu/grids.html}
database, using adopted parameters \teff~=~6300~K, \logg~=~4.0, 
\vmicro~=~1.5~\kmsec, and [Fe/H]~=~$-$2.15.
Detailed discussion of these parameter choices can be found 
in \cite{law13}, but a few points should be highlighted.
First, since \hdeight\ is a nearby ($d$~=~73~$\pm$~3~pc) warm ($B-V$~=~0.36)
main sequence turnoff star, its \teff\ and \logg\ can be very well determined 
from parallax and photometry alone, without reference to spectroscopic
constraints.
Second, \hdeight\ has been analyzed repeatedly for estimation of its
\iso{6}{Li}/\iso{7}{Li} ratio (\citealt{lin13} and references therein), so 
that its atmospheric parameters have been determined many times. 
A list of past analyses can be found in the SAGA database 
\citep{sud08}\footnote{
Available at http://saga.sci.hokudai.ac.jp/wiki/doku.php}; all of
these studies derive similar atmospheric parameters.
Third, the relatively high temperature and gravity of \hdeight\ combine
to produce large densities that favor LTE approximations for atomic level
populations and Planck source functions; such simplifications are much
more problematic for cool metal-poor red giants.
Finally, the same atmospheric conditions lead to continuum opacities that are
dominated by H$^-$ at most wavelengths; continuum scattering is unimportant
even in the $UV$ spectral region.

Abundances were derived following the methods of our previous papers.
We computed synthetic spectra in small spectral intervals surrounding all 
photospheric transitions with the current version of the LTE line analysis 
code MOOG\footnote{
Available at http://www.as.utexas.edu/~chris/moog.html}
\citep{sne73}.
Atomic and molecular line lists for the syntheses were based on the 
transitions compiled in the \cite{kur11} database\footnote{
Available at http://kurucz.harvard.edu/linelists.html.}.
These lists were updated with: 
\textit{(a)} new Fe-group transition probabilities and isotopic/hyperfine
substructures presented in this series of papers;
\textit{(b)} similar data from our earlier publications on $n$-capture 
elements (\citealt{sne09} and references therein); and
\textit{(c)} recent molecular laboratory data on C$_2$ (\citealt{bro13},
\citealt{ram14}), CH \citep{mas14}, CN (\citealt{bro14}, \citealt{sne14}), 
and MgH (\citealt{gha13}, \citealt{hin13}).
The computed solar spectra were smoothed with Gaussians to account for 
instrument profile and solar macroturbulence, and then compared to the 
center-of-disk solar spectrum of \cite{del73} to estimate the 
line-by-line abundances.
This procedure was also followed for \hdeight, except that here we used
the version of MOOG that has scattering included in the continuum source 
function \citep{sob11}. 
As pointed out in earlier papers, for this and other main-sequence 
turnoff stars the scattering contributions are not important.
H$^-$ dominates the continuum opacity, followed by \species{H}{i} in the 
Balmer continuum region (3100$-$3650~\AA).
Therefore we would have obtained essentially the same 
\hdeight\ abundances with the standard MOOG version.

\section{NEW AND RECENT FE-GROUP SOLAR AND \hdeight\ 
         ABUNDANCES\label{newabunds}}

In this section we discuss our analyses of the solar photosphere and \hdeight.
Table~\ref{tab1} contains the abundances, standard deviations, and
numbers of transitions  for neutral and singly-ionized (hereafter, ``ionized'')
 species of these elements.

In Table~\ref{tab2} we compare our ``new'' solar abundances 
to the recent detailed analyses of \cite{sco15} for Sc$-$Ni, and of 
\cite{gre15} for Cu and Zn.
Also tabulated are solar photospheric abundances recommended in the review by 
\cite{asp09}, and the meteoritic abundances recommended by \cite{lod09}.
We quote four values for each species from \citeauthor{sco15} 
and \citeauthor{gre15}.
Three of these abundances come from their papers, and the values that
they obtained from their LTE analyses with the \cite{hol74} model solar
photosphere are from P. Scott (2015, private communication).
Our abundances are essentially identical to their LTE values; differences
in detail can be attributed to line choices, transition probability
assumptions, etc.
\citeauthor{sco15} and \citeauthor{gre15} consider in detail the various 
issues involved in photospheric abundance determinations, including choice 
of model atmosphere grid, 1D versus 3D atmospheric structure, and
local versus non-local thermodynamic equilibrium (LTE versus NLTE) 
in atomic level populations and line source functions.
The reader is urged to consult those papers for additional information on
solar abundance calculations and their uncertainties.

\begin{figure}
\epsscale{1.0}
\plotone{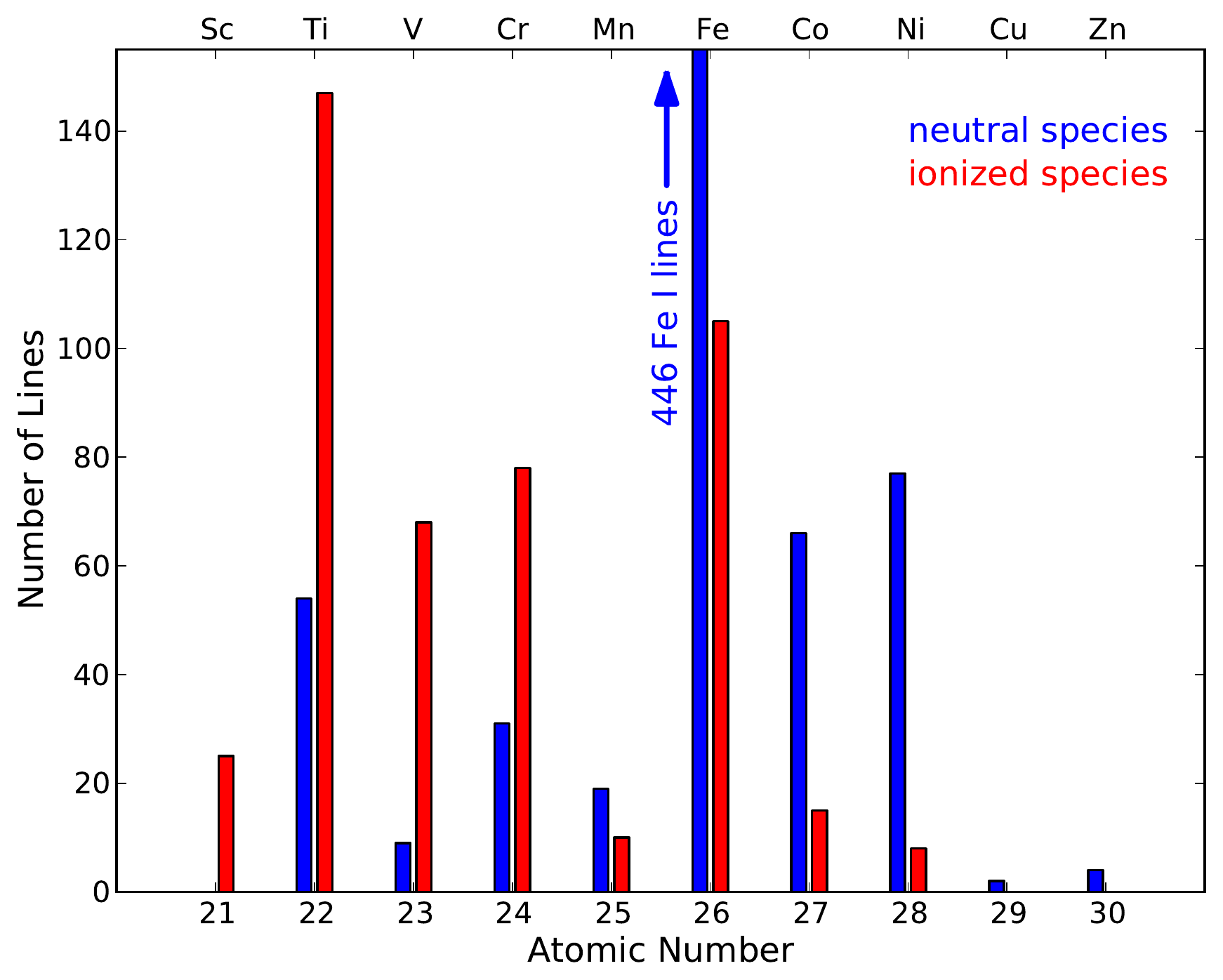}
\caption{
\label{fig1} \footnotesize
The number of measured transitions for neutral
species (blue bars) and ionized species (red bars) in \hdeight.
For display purposes, the vertical extent has been truncated at
150 lines, with an arrow indicating that there are far more \species{Fe}{i}
lines than for other species.
Single bars for Sc, Cu, and Zn indicate that only abundance results 
from \species{Sc}{ii}, \species{Cu}{i}, and \species{Zn}{i} are available. 
}
\end{figure}

Our Fe-group abundances of \hdeight\ have been determined from
a total of 1162 spectral features (551 lines of neutral and
ionized Fe, and 611 of other elements).
The number of lines used for each species is depicted in 
Figure~\ref{fig1}.
Only Cu and Zn abundances have been determined from a single species
with less than five lines each.
For Sc only the ionized lines are detectable, but 25 of these form the
elemental mean abundance.

\begin{figure}
\epsscale{0.9}
\plotone{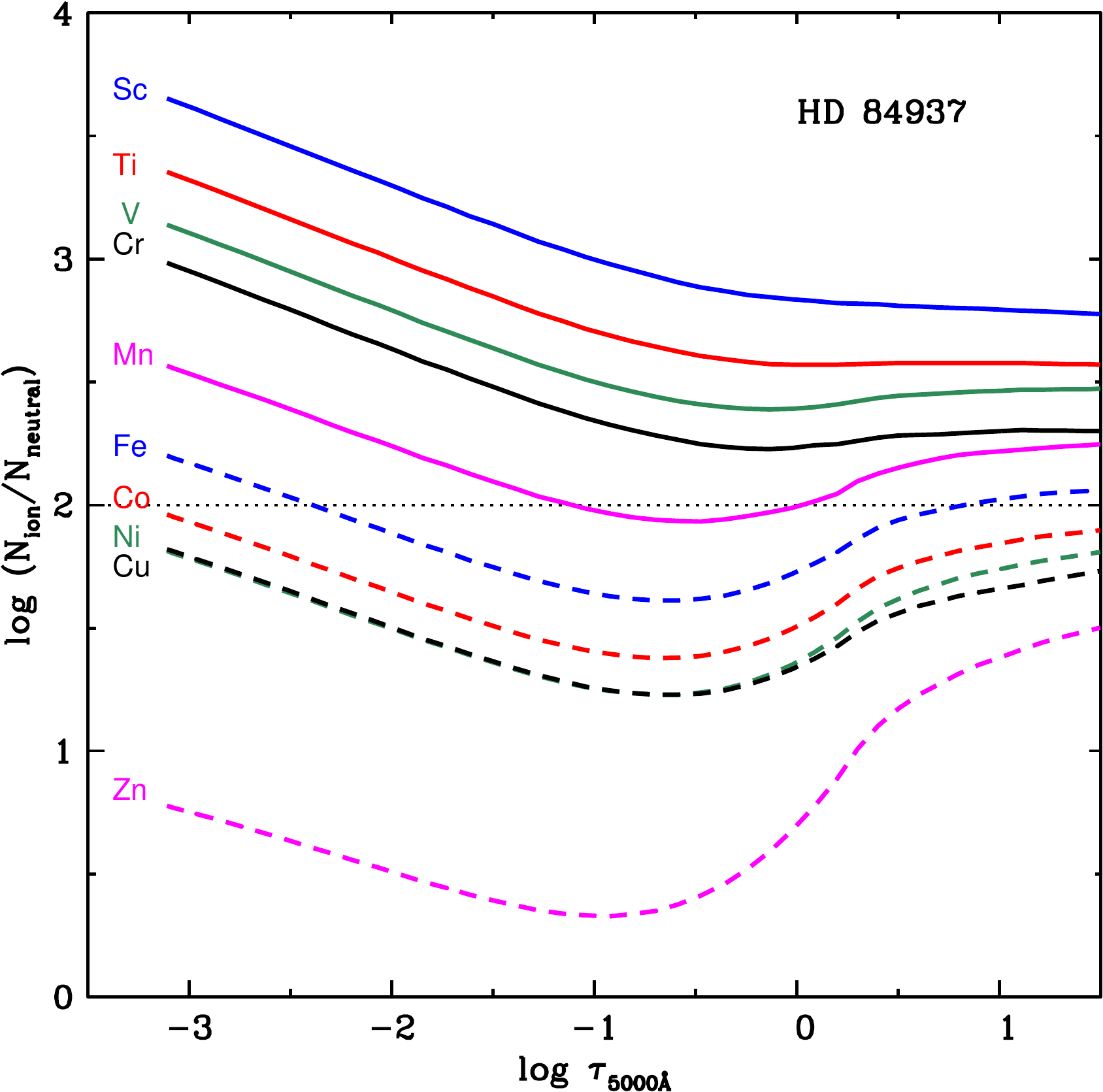}
\caption{
\label{fig2} \footnotesize
Logarithmic number density ionized-to-neutral species ratios
of each Fe-group element as a function of optical depth at 5000~\AA\
in the \hdeight\ atmosphere.
Element labels for the curves are written at the left in the figure.
The dotted line at log(N$_{ion}$/N$_{neutral}$)~=~2 emphasizes the
level at which the Saha ionization balance favors the ion by
a factor of 100.
}
\end{figure}

Eight elements in this study have abundances determined from 
ionized transitions, with no fewer than eight lines forming the 
means of each species abundance.
This is important because ions comprise the vast majority of
the total numbers for all Fe-group elements except Zn.
We illustrate this point in Figure~\ref{fig2} with 
number density ratios log(N$_{ion}$/N$_{neutral}$) plotted as functions 
of optical depth, from straightforward Saha equation calculations.
Five of the elements are completely ionized, defined here as 
log(N$_{ion}$/N$_{neutral}$)~$\gtrsim$~2, throughout the \hdeight\ atmosphere, 
Four other elements have log(N$_{ion}$/N$_{neutral}$)~$\gtrsim$~1.3,
meaning that the ionized species is at least 20 times more abundant
than the neutral species at all atmospheric levels.
The sole exception is Zn, whose ionization potential of 9.39~eV
is high enough to keep a substantial fraction in its neutral species.

Our analyses of Ti, V, Co, and Ni have been discussed in detail in 
previous papers of this series, and those results will be summarized 
together in \S\ref{abtivconi}.
We have comparatively little information on Cu and Zn, and none of
their laboratory data have been generated in this series of papers;
thus we will deal briefly with them together in \S\ref{abcuzn}.
The remaining elements Sc, Cr, Mn, and Fe have transition data published
by the University of Wisconsin laboratory group and collaborators, 
but we have not applied these lab results previously to \hdeight.
Each of the elements will be described in subsections below,
beginning with the fundamental metallicity indicator Fe.

\subsection{Iron\label{abfe}}

Neutral and ionized Fe lines are ubiquitous in the spectra of cool stars,
and this element is almost always used as a surrogate for overall metallicity.
It is unfortunate that \species{Fe}{i} and \species{Fe}{ii} have not 
enjoyed comprehensive laboratory transition data studies in recent years.
However, new accurate $gf$-values for \species{Fe}{i} lines arising from
excited states ($\chi_{lower}$~=~2.4$-$5.1~eV) have been published
by \cite{den14}  and \cite{ruf14}.
If we combine these transition probabilities and those for lines 
with $\chi_{lower}$~$<$~2.4~eV from the earlier study by \cite{obr91}, 
a total of nearly 1100 \species{Fe}{i} lines from one collaborative 
laboratory research group can be applied to stellar spectra.
This combined \species{Fe}{i} list should be on an internally consistent
$gf$ scale, and it has been adopted here.

\begin{figure}
\epsscale{0.9}
\plotone{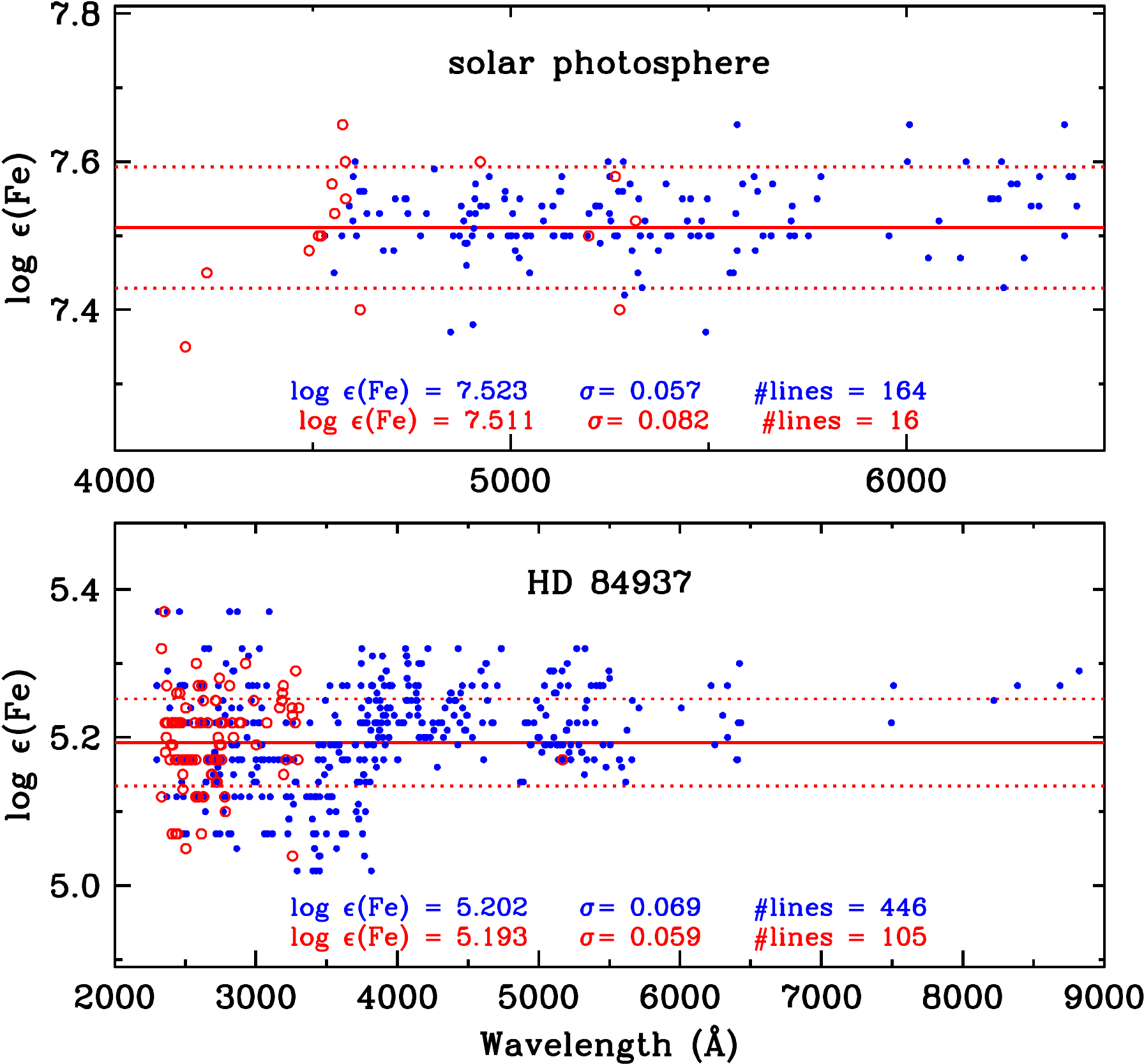}
\caption{
\label{fig3} \footnotesize
Abundances of \species{Fe}{i} and \species{Fe}{ii} lines plotted as 
functions of wavelength for the solar photosphere (top panel) 
and \hdeight\ (bottom panel). 
Blue symbols are for lines of the neutral species and red symbols for the
ionized species.
The species statistics from Table~\ref{tab1} are written at the
bottom of each panel.
A horizontal solid red line denotes the mean abundance from the ionized
species, and the two dotted lines are displaced by $\pm\sigma$
from that mean.
The vertical axis range in each panel is the ionized-species mean $\pm$0.3~dex.
The horizontal axis range is chosen to be large enough to show all line
abundances for the star, and hence will be different for results from
the solar photosphere and for \hdeight.
}
\end{figure}

We first applied the \species{Fe}{i} line list to the solar spectrum, but
restricted the wavelength range to 4500~\AA~$<$~$\lambda$~$<$ 6450~\AA.
The blue limit was set by the steady increases in average \species{Fe}{i} line 
strengths and spectroscopic complexity due to other atomic and
molecular species.  
The number of usable solar features drops precipitously below 4500~\AA.
The red limit was set by the long wavelength end of our \species{Fe}{ii}
line list (to be discussed below); thus the solar Fe abundance derived
here depends on neutral and ionized lines arising in similar 
spectral regions.
In Table~\ref{tab3} we list wavelengths, excitation energies,
log~$gf$ values and their sources, and derived abundances for the 
\species{Fe}{i} lines used in the solar analysis, and in the upper panel 
of Figure~\ref{fig3} we display the photospheric line abundances as a 
function of wavelength.
The scatter ($\sigma$~=~0.057) is small and probably is limited
by the solar analysis uncertainties.
The mean abundance, log~$\epsilon$~=~7.523 (Table~\ref{tab1}),
agrees with the Sc015 value derived with the HM model and
is similar to other solar values (Table~\ref{tab2}).

There are no recent single-source lab transition
studies of \species{Fe}{ii}.
Here we chose to use the $gf$-values from the National
Institute of Standards and Technology (NIST) Atomic Spectra
Database\footnote{http://physics.nist.gov/PhysRefData/ASD/},
which are based on a ``critical compilation'' \citep{fuh06} of 
previously published values.
Since our transition probabilities are taken directly from the NIST website, 
we cite that source for the \species{Fe}{ii} lines in Table~\ref{tab3}.
The original $gf$ data are from (mostly experimental) papers by
\cite{ber96}, \cite{raa98}, \cite{sik99}, \cite{don01}, 
\cite{pic01,pic02}, and \cite{sch04}, which are cited for each line by NIST.

The number of available \species{Fe}{ii} lines for a 
solar abundance analysis is not large, and in the end just 16 were used.
We attempted to analyze only two \species{Fe}{ii} lines with wavelengths
below 4500~\AA, deeming others to be too strong/weak/blended to
yield reliable abundances.
However, the mean abundance derived from \species{Fe}{ii} lines is
in agreement with the \species{Fe}{i} value.
Recognizing the lack of comprehensive lab studies of \species{Fe}{ii},
\cite{mel09} combined some laboratory $gf$ values with ones derived
from a reverse solar analysis.
However, the lower wavelength limit of their study was 4087~\AA. 
Since our goal is to use as many lines as possible down to 2300~\AA\ in the
\hdeight\ spectrum, we have not considered their data further here.

For \hdeight\ we used the complete \species{Fe}{i} list and found 446
useful lines over the wavelength range 
2300~\AA~$\lesssim$~$\lambda$~$\lesssim$ 9000~\AA.
These line abundances are given in Table~\ref{tab3} and plotted
as a function of wavelength in the lower panel of Figure~\ref{fig3}.
This plot shows that the derived abundances have no trend 
with wavelength except for a dip in Fe abundances from lines in the Balmer 
continuum regions, an anomaly that was discussed extensively by 
\cite{roe12} in the study of four metal-poor giants.
The \species{Fe}{i} lines both blueward and redward of this region yield
consistent abundances.
We also derived Fe abundances for the \species{Fe}{i}
lines using the transition probabilities from the Oxford (\citealt{bla95}
and references therein) and Hannover (\citealt{bar91}, \citealt{bar94})
laboratory groups.
For 61 lines in common with the Oxford group,, 
$\langle{\rm log}(gf)_{Oxford}-{\rm log}(gf)_{this study}\rangle$ = $-$0.02 
with $\sigma$ = 0.09. 
With these transition probabilities, the mean \hdeight\ abundance for 
this line subset is 0.06 larger than than our value in Table~\ref{tab1}.
For 67 lines in common with the Hannover group,,
$\langle{\rm log}(gf)_{Hannover}-{\rm log}(gf)_{this study}\rangle$ = $+$0.02 
with $\sigma$ = 0.17.
This large scatter is however caused entirely by disagreements between
our preferred transition probabilities and theirs for vacuum-$UV$ transitions.
Excluding lines with $\lambda$~$<$~3000~\AA, the comparison is much
more favorable:  For the remaining 57 lines 
$\langle{\rm log}(gf)_{Hannover}-{\rm log}(gf)_{this study}\rangle$ = 0.00 with
$\sigma$ = 0.05, and agreement in derived Fe abundance with our value.

Many more \species{Fe}{ii} lines are available in \hdeight\ than in the 
solar photosphere. 
The line abundances (Table~\ref{tab3} and the lower panel
of Figure~\ref{fig3}) have no substantial trends with wavelength.
The derived mean Fe abundances from both species are in agreement.

\subsection{Titanium, Vanadium, Cobalt, and Nickel\label{abtivconi}}

\textit{Titanium:} Our first studies of Fe-group transition data dealt with
\species{Ti}{i} \citep{law13} and \species{Ti}{ii} \citep{woo13}.
Large numbers of useful lines of both species are available in 
solar and \hdeight\ spectra.
The dominant Ti isotope is \iso{48}{Ti}, comprising 73.7\% of the 
total according to the KAERI database\footnote{
Chang, J.: Table of Nuclides, Korea Atomic Energy Research Institute
(KAERI); available at: http://atom.kaeri.re.kr/ton/.}.
The remaining 26.3\% is spread fairly evenly among the 
\iso{46,47,49,50}{Ti} isotopes.
Fortunately, the isotopic splitting, and hyperfine structures 
(hereafter, hfs) of odd-N isotopes \iso{47,49}{Ti}, are very 
narrow, producing negligible broadening of Ti emission lines in very 
high-resolution laboratory spectra.
Therefore \citeauthor{law13} and \citeauthor{woo13} treated the
Ti transitions as single lines. 
We quote their results in Table~\ref{tab1}.
For the solar photosphere the line-to-line scatter, $\sigma$~$\lesssim$~0.05,
is consistent with uncertainties in extracting abundances from matching
observed and synthetic spectra.

For \hdeight\ \cite{law13} noted a dip of $\sim$0.2~dex in 
derived \species{Ti}{i} and \species{Fe}{i} abundances in the near-$UV$, 
$\lambda$~$\sim$ 3100$-$3700~\AA.
This abundance depression amounts to $\sim$0.15~dex on average, about half
that found by \cite{roe12} in their four giant stars.
\cite{woo13} found a similar anomaly for \species{Ti}{ii} lines but mostly
confined to those with excitation energies $\chi$~$>$~0.6~eV (see their
Figure 12).
That paper's \S8 discusses this issue, arguing that the abundance dip 
for \hdeight\ is connected to the Balmer continuum opacity.
For this warm main-sequence turnoff star H$^-$ dominates all opacity 
sources except in the 3100$-$3700~\AA\ region, and the abundance dip is 
more mild than in cooler, lower gravity stars with weaker H$^-$ opacity.
Here we note that in Table~\ref{tab1} we have adopted the mean of 
all \species{Ti}{ii} line abundances. 
If we were to neglect the 3100$-$3700~\AA\ lines, the mean Ti abundance
from this species would rise to log~$\epsilon$~$\simeq$ 3.12, in better
agreement with the value derived from \species{Ti}{i} lines.
For our purposes this is not a concern, but future studies with a more
comprehensive physical analysis will be welcome.

\textit{Vanadium:} We presented new laboratory transition data for 
\species{V}{i} in \cite{law14} and for \species{V}{ii} in \cite{woo14a}.
Vanadium's sole naturally-occurring isotope is \iso{59}{V}.
Substantial hfs broadening is present in spectral features of
both \species{V}{i} and \species{V}{ii}.
Laboratory hfs data are available for the majority of astronomically-relevant 
transitions, and they were taken into account when possible 
(\citeauthor{law14}, Table~5, Figure~6; 
\citeauthor{woo14a}, Table~5, Figure~4); see those papers for detailed
hfs discussions.
Good agreement was found between our solar abundances from both V species and 
the recommended photospheric and meteoritic values (Table~\ref{tab2}).
The Sco15 results for \species{V}{i} are about 0.05 dex smaller, and
\species{V}{ii} was not considered by those authors.
In HD~84937 the \species{V}{i} lines are all extremely weak, and the
dominant analytical problem becomes one of detection (\citeauthor{law14}, 
Figure 7).
Nonetheless, satisfactory abundance agreement was achieved between the
two V species in this star.

\textit{Cobalt:} A new lab/stellar study of \species{Co}{i} has been 
completed by \cite{law15}.  
This element's only naturally-occurring isotope is \iso{59}{Co}, and
most \species{Co}{i} transitions have significant hfs.
For the solar photosphere, synthetic spectrum computations for 82 lines, 
most of which have hfs components included in the computations, 
yielded the mean abundance quoted in Table~\ref{tab1}.
There are 23 lines of \species{Co}{ii} identified by \cite{moo66} in the
3300$-$4700~\AA\ spectral range, but these lines have not had recent
transition probability analyses.
The NIST database gives poor grades to their $gf$-values, thus \cite{law15}
did not investigate this species in the Sun.
Application of the new \species{Co}{i} lab data to the \hdeight\ spectrum 
led to the abundance given in Table~\ref{tab1}. 
Fortunately, the HST/STIS spectrum contains lines of \species{Co}{ii} with
more reliable transition probabilities.
A total of 15 lines of this species yielded a mean abundance that is
in excellent agreement with that obtained from \species{Co}{i}.

\textit{Nickel:} \cite{woo14b} reported new transition probabilities 
for \species{Ni}{i}.
While Ni lines have negligible hfs,  in the red spectral region the 
isotopic splitting among its five 
naturally-occurring isotopes (\iso{58}{Ni}, 68.1\%, KAERI database; 
\iso{60}{Ni}, 26.2\%; \iso{61}{Ni}, 1.1\%; \iso{62}{Ni}, 3.6\%; 
and \iso{64}{Ni}, 0.9\%) is detectable in the solar spectrum.
\citeauthor{woo14b} included isotopic substructure when possible in 
their calculations.
For the solar photosphere their derived abundance was about 0.06~dex larger
than other values from the literature (Table~\ref{tab2};
see \citeauthor{woo14b}), but within the mutual uncertainties of their
and other values.
Transitions of \species{Ni}{ii} have been identified in the solar spectrum
\citep{moo66} but they arise from excited-state lower levels 
($\chi$~$\gtrsim$ 3~eV), and their transition probabilities have not
been studied in the lab for about three decades.
Lines of \species{Ni}{i} redward of 5000~\AA\ are numerous in the 
solar spectrum but are mostly undetectable in \hdeight.
But many transitions of this species at shorter wavelengths are available 
for analysis in this very metal-poor star. 
Moreover, strong lines of \species{Ni}{ii} from lower excitation levels 
($\chi$ = 1.0$-$2.0~eV) with more recent lab $gf$-values \citep{fed99}
can be found in the vacuum $UV$ spectral range.
In Table~\ref{tab1} we quote the \citeauthor{woo14b} abundances
from both Ni species in \hdeight,  which clearly are in agreement.

\subsection{Scandium\label{abcs}}

For this paper we only consider \species{Sc}{ii} transitions.
All useful solar \species{Sc}{i} lines are very weak. 
In the photospheric abundance studies of \cite{neu93} and Sco15
the strongest line of this species is at 5671.8~\AA, 
with EW~$\simeq$~12~m\AA, or 
log~$RW$~$\equiv$~log($EW/\lambda$)~$\simeq$~$-$5.7.
Most other useful solar lines of this species are much weaker.
The low metallicity of \hdeight\ combined with its very large atmospheric 
N$_{\rm Sc II}$/N$_{\rm Sc I}$ ratio 
(Figure~\ref{fig2}) conspire to make its \species{Sc}{i} lines undetectable.

The most recent laboratory transition probabilities are from \cite{law89},
and they are adopted here for all \species{Sc}{ii} transitions.
When possible we included hfs in the abundance calculations.
We calculated the hfs patterns from hyperfine constants given in
\cite{arn82}, \cite{you88}, \cite{man89a}, \cite{man89b}, and \cite{vil92}.

\begin{figure}
\epsscale{1.0}
\plotone{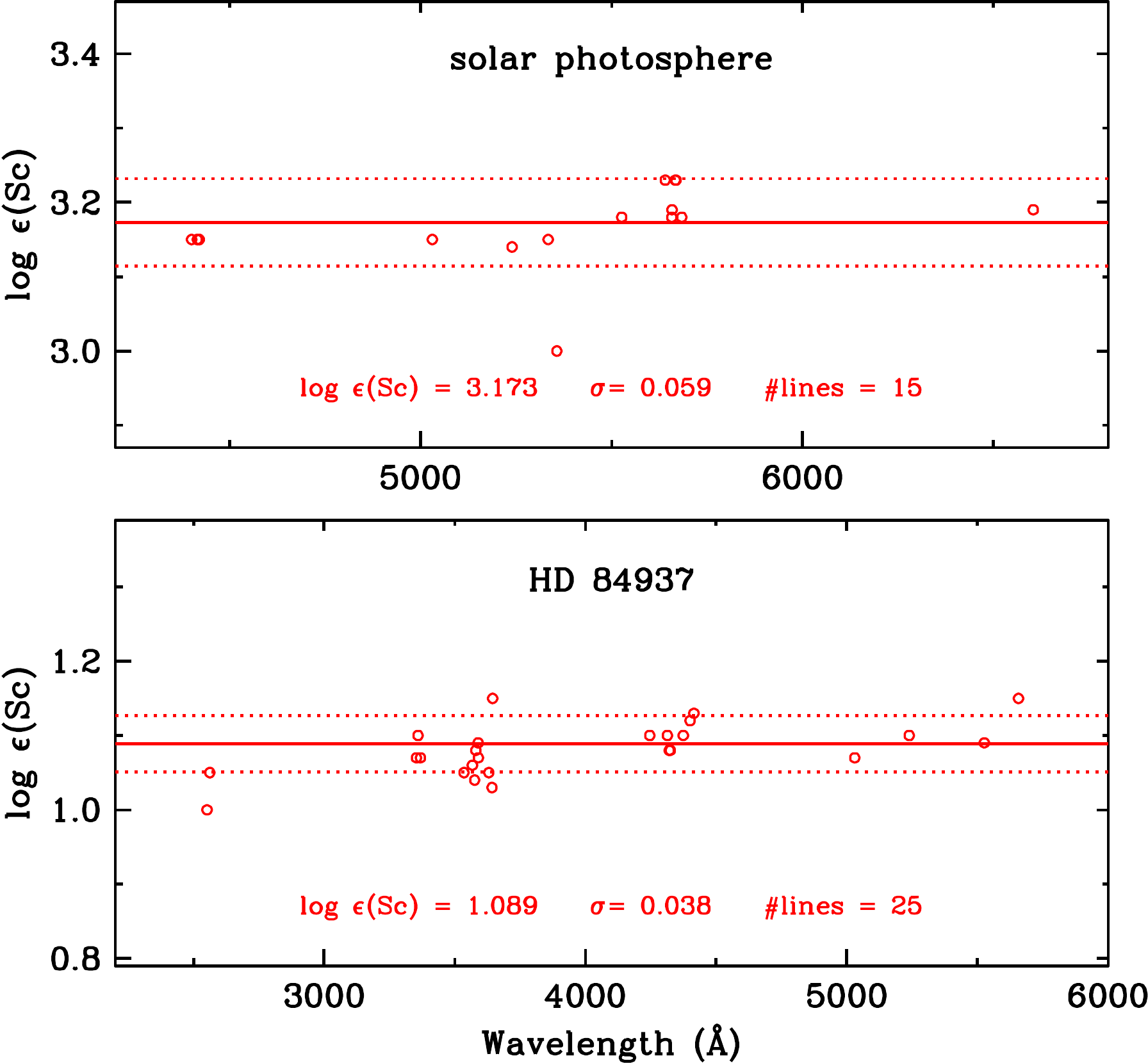}
\caption{
\label{fig4} \footnotesize
Abundances of \species{Sc}{ii} lines plotted as functions of
wavelength.
All lines and points are as defined in Figure~\ref{fig3}.
}
\end{figure}

The majority of useful solar photospheric \species{Sc}{ii} lines are in 
the yellow-red, and from 15 of them (Table~\ref{tab4})
we derive a Sc abundance of log~$\epsilon$~=~3.17
(Table~\ref{tab1} and Figure~\ref{fig4} ) 
that is in good agreement with other
photospheric values (Table~\ref{tab2}).\footnote{
The five Sc II lines between 5600 and 5700~\AA\ are part of the z$^3$P 
to a$^3$P multiplet.   
This multiplet is primarily responsible for the apparent Saha imbalance of 
Sc in the Sun (\eg, \citealt{neu93}).  
Branching fraction measurements on this minor multiplet are challenging 
due to its large wavelength separation from the  dominant UV branches of 
the z$^3$P upper levels and absence of intermediate wavelength lines.  
Re-measurement of \species{Sc}{i} and \species{Sc}{ii} branching fractions 
is underway, and more than 90\% of the \cite{law89} log($gf$) values have 
been reproduced to within error bars.  
Preliminary evidence suggest that the log($gf$) values of the z$^3$P 
to a$^3$P multiplet by \cite{law89} may be too small by  about twice 
their $\pm$0.05~dex uncertainty.}
The photospheric Sc abundance is about 0.1~dex higher than the
meteoritic abundance recommended by \cite{lod09}, a discrepancy already
noted by Sco15.
Resolution of this discrepancy is beyond the scope of this work.

More \species{Sc}{ii} violet-$UV$ transitions become available in the \hdeight\ 
spectrum, and the 25 lines used in our analysis (Table~\ref{tab4})
have excellent internal abundance agreement, with $\sigma$~=~0.038
(Table~\ref{tab1} and Figure~\ref{fig4}).

\subsection{Chromium\label{abcr}}

This element has received recent attention from laboratory spectroscopists:
\species{Cr}{i} by \cite{sob07}, and \species{Cr}{ii} by 
\cite{nil06}, \cite{gur10}, and \cite{eng14}.
Unfortunately this has not led to agreement in abundances derived from
the two species.
\cite{mcw95} discovered that abundances derived from lines of 
\species{Cr}{i} exhibit a steady decline in [Cr/Fe] at metallicities 
[Fe/H]~$\lesssim$~$-$2, reaching [Cr/Fe]~$\sim$~$-$0.5 at 
[Fe/H]~$\lesssim$~$-$3.
This trend has been confirmed and refined by subsequent 
surveys, \eg, \cite{cay04}; see a summary in Figure~20 of \cite{kob06}.
However, the less frequent results from \species{Cr}{ii} analyses find no 
obvious decline: [Cr/Fe]~$\simeq$~0 over the entire metallicity range 
$-$3.4~$\lesssim$~[Fe/H]~$\lesssim$~$+$0.4 (as summarized 
by \citealt{kob06}). 
Indeed, \cite{roe14} suggest that \species{Cr}{ii} lines yield
Cr overabundances in low metallicity stars: [Cr/Fe]~$\simeq$~$+$0.2.

Cr has four naturally-occurring isotopes: \iso{50}{Cr}, 4.2\%;
\iso{52}{Cr}, 83.8\%; \iso{53}{Cr}, 9.5\%;
and \iso{54}{Cr}, 2.4\% (KAERI database).  
Not only does \iso{52}{Cr} dominate the elemental abundance, but isotope
shifts are small. 
We inspected very high-resolution Cr emission-line lab spectra obtained with 
the National Solar Observatory (NSO) 1~m FTS \citep{bra76} and accessible
in the NSO archive\footnote{
available at  http://diglib.nso.edu/}, and found all line profiles to
be narrow and symmetric.
Therefore we ignored hfs and isotopic substructure in our abundance 
analyses.  

\begin{figure}
\epsscale{1.0}
\plotone{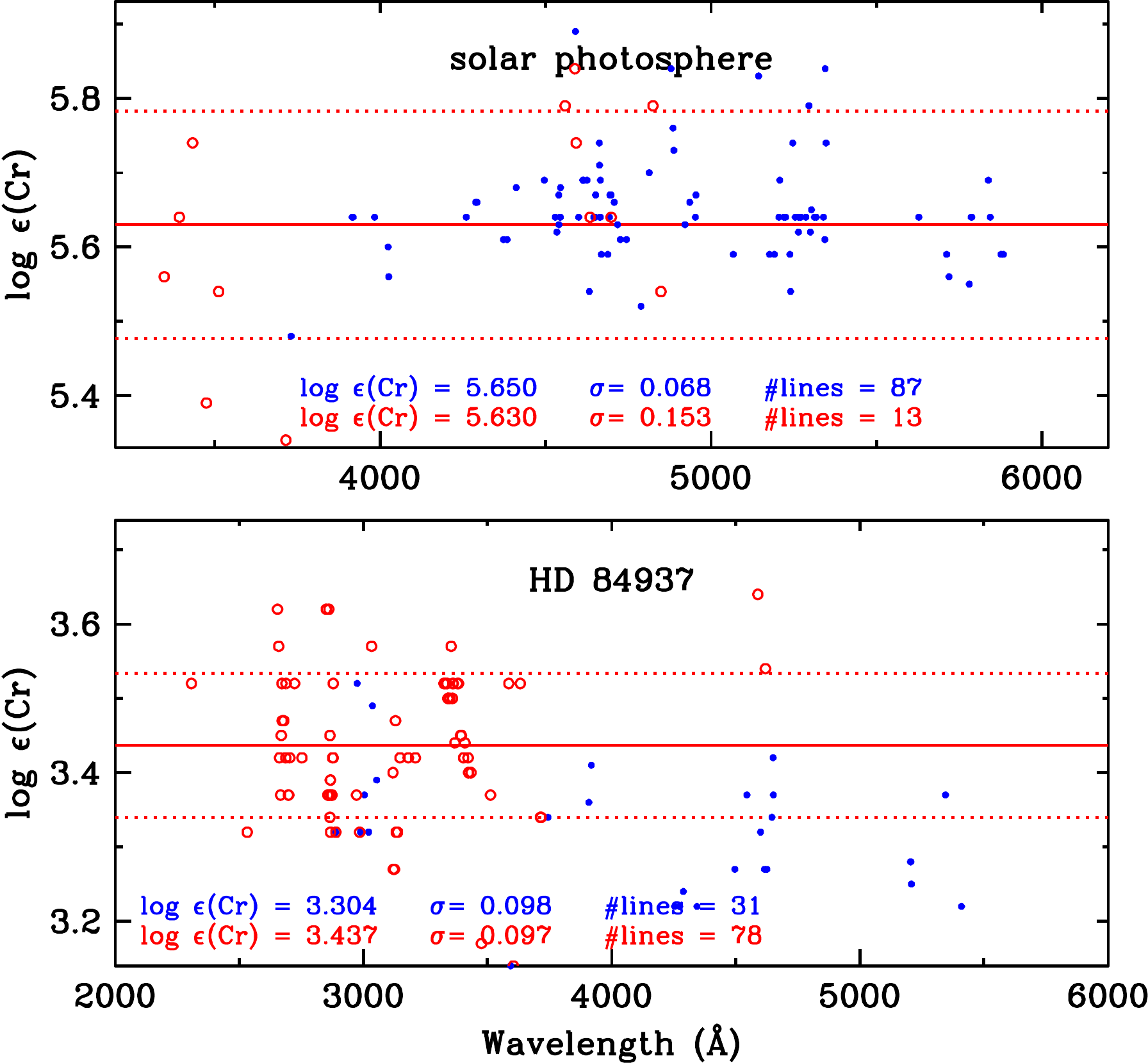}
\caption{
\label{fig5} \footnotesize
Abundances of \species{Cr}{i} and \species{Cr}{ii} lines plotted as functions of
wavelength.
All lines and points are as defined in Figure~\ref{fig3}.
}
\end{figure}

The line abundances for \species{Cr}{i} and \species{Cr}{ii} are listed in
Table~\ref{tab4}.
There are many available transitions for both species in the Sun and
\hdeight, but inspection of Figure~\ref{fig5} reveals that scatters
in the Cr line abundances of the Sun and \hdeight\ are larger than for 
other Fe-group elements.
For the Sun, the abundance means (Table~\ref{tab1}) derived from 
neutral and ionized lines are in agreement. 
A handful of \species{Cr}{i} lines yield anomalously large
abundances, but these do not stand out in plots of abundance versus
excitation potential or line strength.
Very low values for a couple of near-$UV$ \species{Cr}{ii} lines are
mainly responsible for the large scatter ($\sigma$~=~0.153) in its
mean abundance; neglecting these lines would bring the two species
abundances into better agreement.

For \hdeight, again there is large scatter in individual line
abundances (Table~\ref{tab1}, Figure~\ref{fig5}), 
and the now-familiar 
offset between \species{Cr}{i} and \species{Cr}{ii} appears.  
However, the abundance difference is not large, 0.13~dex, and
is at the level seen for this metallicity regime in Figures~20 and 21
of \cite{kob06}.
Note that three \species{Cr}{i} lines yield abundances
about 0.2~dex less than the mean value for this species, but excluding 
them would not change the mean significantly.
\cite{sco15} have questioned the accuracy of
the \cite{nil06} \species{Cr}{ii} transition probabilities for those 
lines involved in their solar abundance determinations.
To test whether the chosen transition probabilities
for \species{Cr}{ii} lines could be contributing to the
large line-to-line scatter and the ionization mismatch with abundances
derived from \species{Cr}{i}, we recalculated the \species{Cr}{ii} 
abundances with the $gf$ values posted to the
\cite{kur11} database prior to \cite{nil06}.
The mean \hdeight\ abundance from \species{Cr}{ii} decreased by 0.13~dex,
cancelling the ionization balance problem.
However, the internal scatter for \species{Cr}{ii} grew substantially
to $\sigma$~=~0.25, an unacceptably large value.
This issue should be explored in more detail as part of future Cr 
studies in metal-poor stars; renewed lab studies should be
part of that effort.

\subsection{Manganese\label{abmn}}

Relative deficiencies of Mn in low metallicity stars have been known
since the early days of high-resolution stellar spectroscopy.
This anomaly was noted by \cite{wal63}, and many studies 
suggest that [Mn/Fe]~$\lesssim$~$-$0.4 for [Fe/H]~$<$~$-$1 
(see Figure~22 of \citealt{kob06}).
Most of the Mn abundance analyses in the literature have employed only
\species{Mn}{i} lines.
This is risky because Mn is almost completely ionized in \hdeight\ 
(Figure~\ref{fig2}) and in the Sun as well.

\begin{figure}
\epsscale{1.0}
\plotone{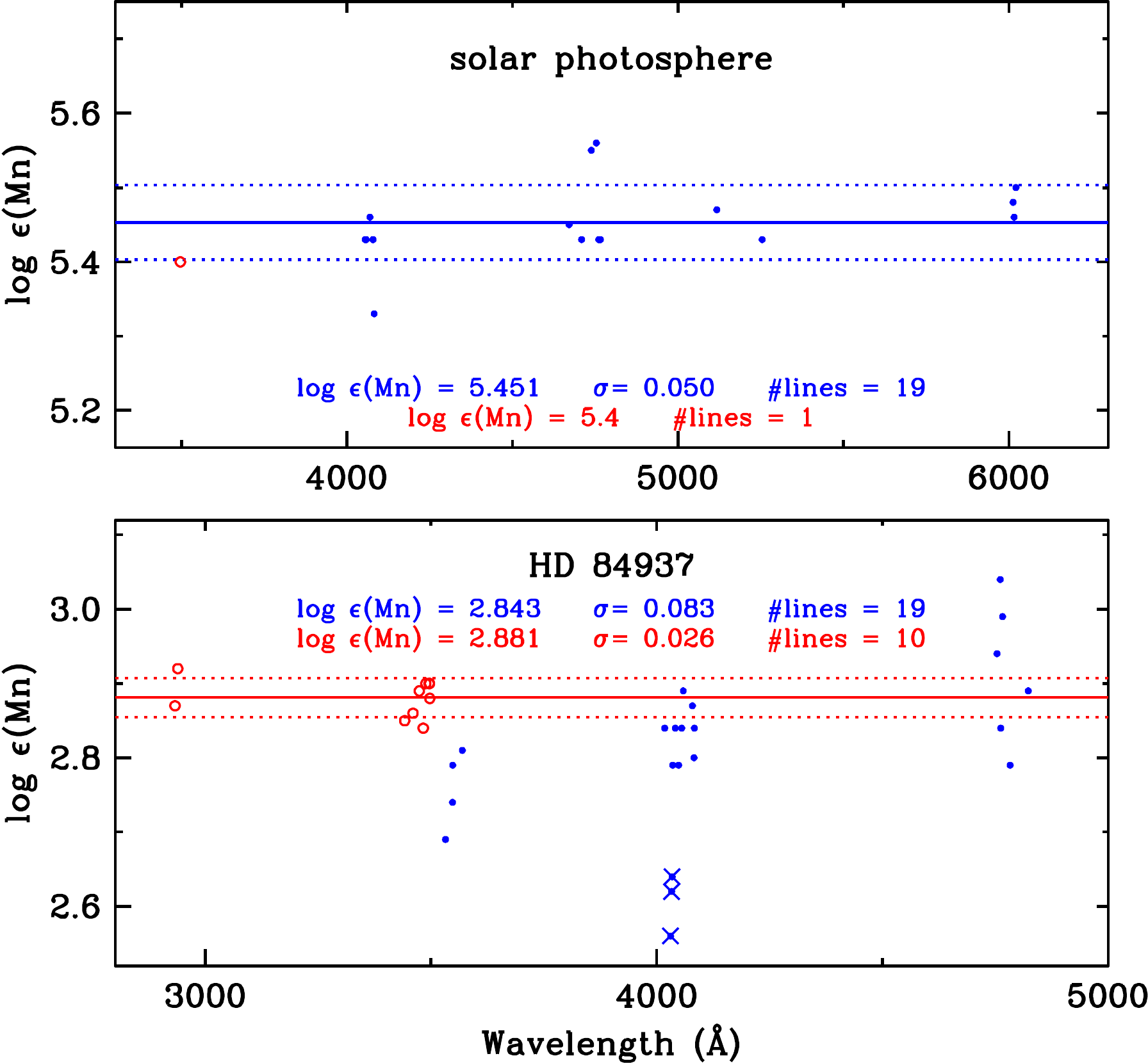}
\caption{
\label{fig6} \footnotesize
Abundances of \species{Mn}{i} and \species{Mn}{ii} lines plotted as functions of
wavelength.
Most lines and points are as defined in Figure~\ref{fig3}.
In the top panel, we use blue horizontal lines to indicate that the mean 
and standard deviations are from \species{Mn}{i}, because there is only one 
\species{Mn}{ii} abundance point, and it is not very well determined.
In the lower panel we use $\times$ symbols to denote the abundances derived
for the \species{Mn}{i} resonance triplet.
Their abundances are depressed relative to other \species{Mn}{i} transitions, and
hence they are not included in the abundance statistics.
}
\end{figure}

Laboratory investigations of \species{Mn}{i} and \species{Mn}{ii} have
been published by \cite{den11}. 
Here we adopt their $gf$ values and hfs patterns exclusively, so all of the
transition data comes from a single source.
The solar photospheric Mn abundance derived from \species{Mn}{i} lines
appears to be trustworthy. 
The line-to-line scatter is small (Table~\ref{tab4}, 
Figure~\ref{fig6}, 
upper panel), and the mean abundance agrees with other solar abundance 
estimates within mutual uncertainties (Table~\ref{tab2}).
We searched for suitable solar \species{Mn}{ii} lines from the 
\citeauthor{den11} list but found only one, 3497.5~\AA.
Unfortunately, that transition is part of a clump of other Fe-group lines. 
We were only able to derive an approximate abundance,
log~$\epsilon$~$\simeq$~5.4 for this line.
It is consistent with the abundance derived from \species{Mn}{i} lines,
but should be viewed with caution.
We have chosen not to quote this unreliable estimate in
Table~\ref{tab1}, and it is not considered further in the paper.

For \hdeight\ there is significant scatter among the \species{Mn}{i} line
abundances (Table~\ref{tab4}, Figure~\ref{fig6}, 
lower panel).
In particular, the 4030, 4033, 4034~\AA\ resonance triplet abundances
are $\sim$0.3~dex lower than the means of other Mn neutral and ionized
species lines.
We suggest that this is further evidence for departures from LTE in 
\species{Mn}{i} line formation involving ground-state lines.
\cite{ber08} computed such NLTE correction factors (see their Table~2),
and found values to range from $\simeq$0.0~dex at solar metallicity to as
much as $\simeq$0.7~dex at [Fe/H]~=~$-$3.
In the \teff/\logg/[Fe/H] domain of \hdeight\ the proposed NLTE corrections
for \species{Mn}{i} lines are typically $\simeq$0.3$-$0.4~dex, but 
for the resonance triplet they are $\simeq$0.5$-$0.6~dex.
This differential correction would nearly erase the abundance mismatch we
see in the triplet lines compared to the rest of this species.
Independent computation of NLTE effects is beyond the scope of this
paper.
Therefore we have chosen to exclude the \species{Mn}{i} triplet lines from 
the species mean given in Table~\ref{tab1} and shown in 
Figure~\ref{fig6}.
If we include the resonance triplet in the \species{Mn}{i} abundance 
statistics, the change would be small:  
$\langle$log $\epsilon\rangle$~=~2.811~$\pm$~0.024 .
The mean Mn abundance derived from \species{Mn}{ii} lines in \hdeight\ 
(Table~\ref{tab1}) is only 0.04~dex larger than that derived from
\species{Mn}{i} without the triplet.

\subsection{Copper and Zinc\label{abcuzn}}

Copper and zinc present only a small number of detectable transitions of their
neutral species in the spectra of the Sun and \hdeight.
Strong lines of their first ions occur only in the deep $UV$ spectral
region, shortward of 2300~\AA.
We thus were unable to assess ionization equilibrium for either of these
elements.

Both hyperfine and isotopic effects split \species{Cu}{i} lines into
multiple components.
There are two stable isotopes: \iso{63}{Cu}, 69.2\%; and \iso{65}{Cu}, 
30.8\% (KAERI database).
We adopted these percentages for the solar photosphere and, lacking 
isotopic information, for \hdeight. 
The hyperfine components were adopted from the \cite{kur11} database.
Accurate transition probabilities are known for the lines of interest
here, and we adopted the values given in the NIST database.
For the Sun, line-to-line agreement is excellent and the 
mean Cu abundance is consistent with the Sco15 LTE value computed with the 
\cite{hol74} atmosphere, and with \cite{asp09} recommended 
photospheric abundance.
Only the two \species{Cu}{i} resonance lines at 3247.5 and 3274.0~\AA\
could be detected in our \hdeight\ spectra, and they yield Cu abundances
in fortuitously perfect agreement (Table~\ref{tab4}).
Here we have arbitrarily assigned an uncertainty $\sigma$~=~0.07, a 
typical value for the other abundances in \hdeight.

Zinc is the only Fe-group element with a substantial population in the
neutral species (Figure~\ref{fig2}).
Its detectable transitions in our spectra arise from excited lower
energy states, $\chi$~=~4.0~eV.
Both Saha and Boltzmann factors matter in the \species{Zn}{i} level
populations.  
We were able to derive mean Zn abundances from three lines in the
solar photosphere and four lines in \hdeight.
The line-to-line scatters are acceptable, and in the case 
of the Sun our derived abundance of log~$\epsilon$(Zn)~=~4.61 splits the 
difference between other photospheric ($\simeq$4.55) and meteoritic 
(4.65) values (Table~\ref{tab2}).
However, for \hdeight\ our abundance is not as well constrained as
are those of other Fe-group elements, and should be viewed with caution.

\subsection{Abundance Uncertainties and Comments\label{abcomm}}

Inspection of the \hdeight\ abundances in Table~\ref{tab1} suggests
that those derived from neutral and ionized species agree very well.
In no case do the differences exceed their 1$\sigma$ uncertainties.
Thus we believe that, at least for seven of the ten Fe-group elements,
Saha ionization balance is achieved with our standard
spectroscopic analysis.
Neutral species yield the same elemental abundances as do the ionized species.
The derived abundances are the result of real nucleosynthesis output,
not just stellar atmospheric effects.

This result holds with reasonable atmospheric
parameter excursions from our baseline model of \teff~=~6300~K, \logg~=~4.0,
\vmicro~=~1.5~\kmsec, and [Fe/H]~=~$-$2.15.
The scatter in these parameters reported in the literature for \hdeight\ are 
$\sigma$(\teff)~$\simeq$~80~K, $\sigma$(\logg)~$\simeq$~0.15,
$\sigma$([M/H])~$\simeq$~0.15, and $\sigma$(\vmicro)~$\simeq$~0.1
(from papers included in the SAGA database).
We calculated abundances for representative lines of each species for 
different atmosphere models that go beyond these probable uncertainties.
In general, Fe-group neutral species transitions collectively respond 
in a common way to parameter changes, as do all of the Fe-group ions.
For temperature changes $\Delta$\teff~=~$\pm$150~K, 
$\Delta$(log~$\epsilon$(\species{X}{i}))~$\simeq$ $\pm$0.16, and 
$\Delta$(log~$\epsilon$(\species{X}{ii}))~$\simeq$ $\pm$0.06.
The positive correlation with \teff\ is due to H$^-$ continuum opacity
changes, and the difference between neutrals and ions is due to their
own Saha balances.
For gravity changes $\Delta$\logg~=~$\pm$0.25,
$\Delta$(log~$\epsilon$(\species{X}{i}))~$\simeq$ $\mp$0.06, and
$\Delta$(log~$\epsilon$(\species{X}{ii}))~$\simeq$ $\pm$0.04; this
is a direct Saha effect.
For model metallicity changes $\Delta$[M/H]=~$\pm$0.25,
$\Delta$(log~$\epsilon$(\species{X}{i}))~$\simeq$ $\pm$0.01, and
$\Delta$(log~$\epsilon$(\species{X}{ii}))~$\simeq$ $\pm$0.01.
This essentially null sensitivity to metallicity is because 
the major electron donor for the continuum opacity is
H itself; heavy elements are minor contributors.
Finally, since \hdeight\ is a weak-lined star the microturbulent
velocity sensitivity is not large.
For $\Delta$\vmicro~=~$\pm$0.15, if log~$RW$~$\lesssim$~$-$5.5 then
$\Delta$(log~$\epsilon$(X))~$\simeq$ 0.00
for lines of either neutral or ionized species, 
and if log~$RW$~$\lesssim$~$-$4.5 then
$\Delta$(log~$\epsilon$(X))~$\simeq$ $\mp$0.05.

However, these abundance sensitivities are muted in 
relative abundance ratios.
In our discussion to follow, we will concentrate on 
relative abundances [X/Fe].
We will be comparing neutral to neutral and ion to ion, and in almost all cases 
$\Delta$(log~$\epsilon$(\species{X}{i}/\species{Fe}{i}) and/or
$\Delta$(log~$\epsilon$(\species{X}{ii}/\species{Fe}{ii}) $\lesssim$ $\pm$0.02.
Within our standard abundance analysis approach, model atmosphere parameter
uncertainties cannot substantially perturb the abundances shown in
Table~\ref{tab1}.

Several cautions should be kept in mind in interpreting our results.
\textit{(1)} There are three ``missing'' species 
(\species{Sc}{i}, \species{Cu}{ii}, \species{Zn}{ii}), and three either 
with no solar abundance or an untrustworthy value 
(\species{Mn}{ii},\species{Co}{ii}, \species{Ni}{ii}).
\textit{(2)} Line formation in the Balmer continuum wavelength region is
inadequately described by our standard modeling techniques.
\textit{(3)} Internal line-to-line scatters are large for \species{Cr}{i}
and \species{Cr}{ii}, and the agreement between the mean abundances for
these two species is not as good as for the other Fe-group elements.
\textit{(4)} \species{Mn}{i} line-to-line scatter is large, and the
resonance triplet yields abundances in \hdeight\ that are nearly a factor
of two lower than the mean of other neutral and ionized lines of this
element.
\textit{(5)} Departures from LTE, not discussed in detail
here, have been subjected to detailed calculations by others.
In general, the neutral species appear to be much more sensitive to
NLTE effects, leading to small predicted changes in derived solar
abundances ($\lesssim$0.1~dex, \eg, \citealt{gre15} and references therein),
but perhaps much larger corrections at low metallicities ($\lesssim$0.5~dex,
\eg, \citealt{lin12} and references therein).
Collectively, these issues clearly indicate the need for more laboratory
effort, particularly for ionized species, and for more stellar
atmosphere and line formation modeling effort.
Our results have yielded promising results, but much further work
in these areas will be welcome.


\section{ABUNDANCE CORRELATIONS AMONG SC, TI, AND V IN 
         \hdeight\ AND OTHER LOW-METALLICITY STARS\label{results}}

\begin{figure}
\epsscale{1.0}
\plotone{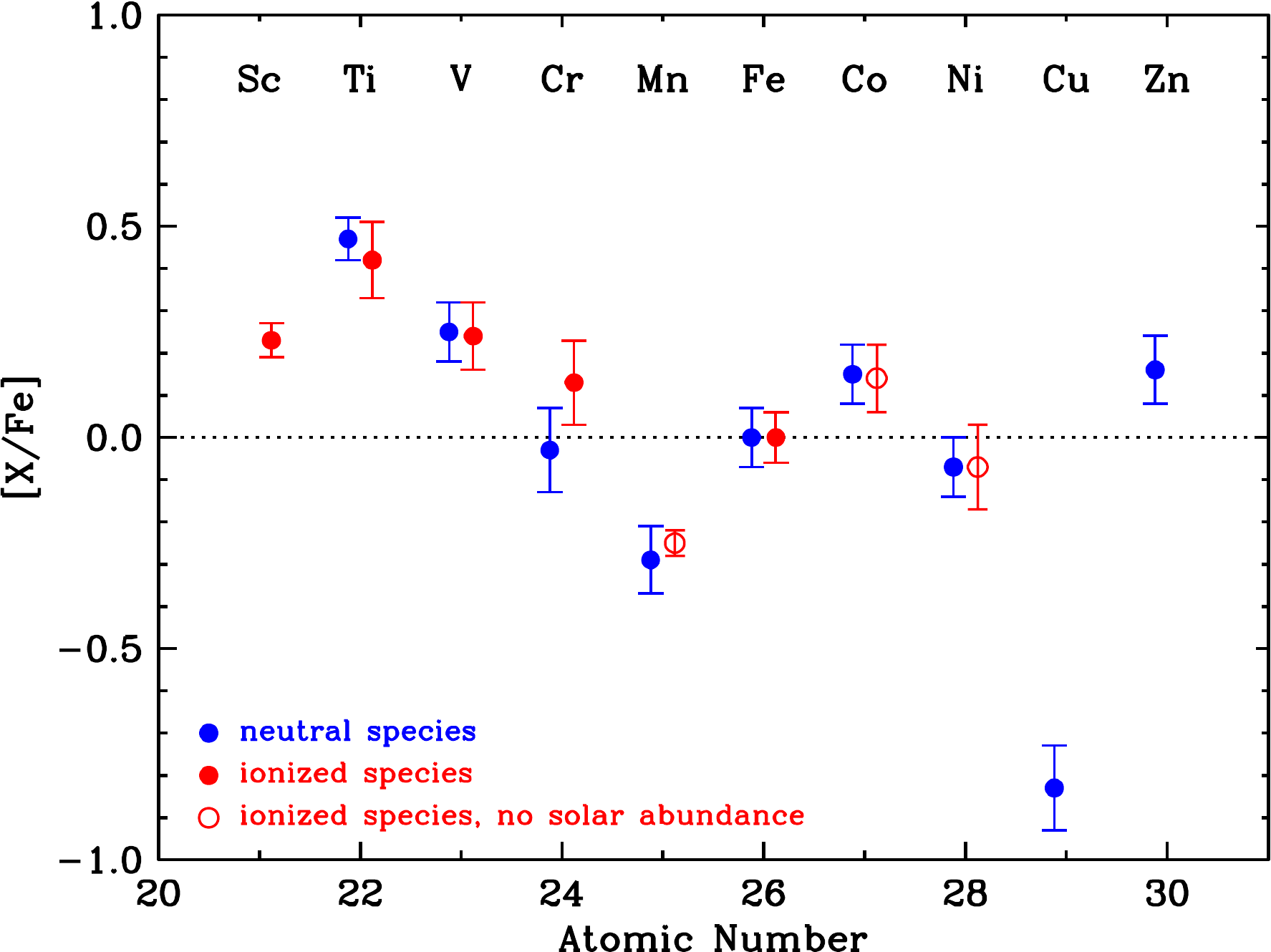}
\caption{
\label{fig7} \footnotesize
Abundance ratios of all species with respect to Fe as a function
of atomic number.
For each element, neutral species abundances are calculated  with respect 
to the \species{Fe}{i} abundance and plotted in blue symbols offset to the left, 
and ions are calculated with respect \species{Fe}{ii} and are in red offset to 
the right.
The error bars are $\pm 1\sigma$ values from Table~\ref{tab1}.
}
\end{figure}

In Figure~\ref{fig7}  and Table~\ref{tab5}
we present a summary of our results for the relative abundances of 
iron-group elements in \hdeight, using the traditional square bracket units.
Our compilation includes both neutral and ionized values
where both are available. 
In Figure~\ref{fig7} abundances determined from neutral species with respect
to \species{Fe}{i} are indicated by blue filled circles and from ionized 
species with respect to \species{Fe}{ii} by red circles. 
As noted above there are a few elements $-$ denoted by open red circles $-$
where no ionized solar values are available.
In those cases solar ionization balance was assumed to determine the 
relative \hdeight\ abundances, \ie, 
log~$\epsilon$(\species{X}{ii})$_\odot$~$\equiv$ 
log~$\epsilon$(\species{X}{i})$_\odot$.
It is clear from the figure that there is good agreement between the neutral
and ionized species for the observed abundances in \hdeight.
We also have listed overall mean relative abundances for these elements in
Table~\ref{tab5}, where we have averaged the neutral and
ionized abundances (where available).
The [X/Fe] values for \hdeight\ in this table will be used in all further
discussion in this paper.

\begin{figure}
\epsscale{1.3}
\includegraphics[bb=30 30 600 600,angle=0,scale=2.0,width=6.7in]{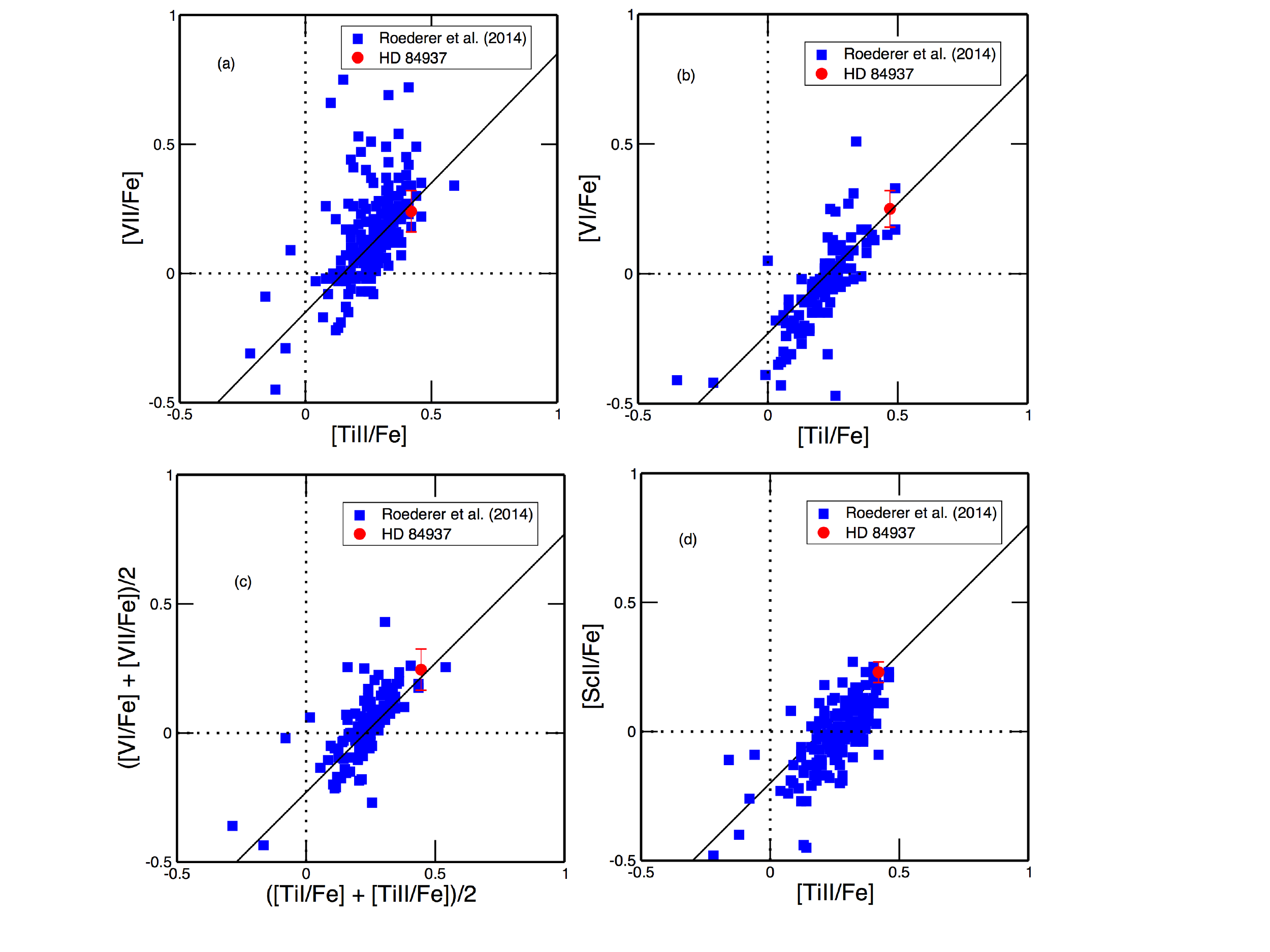}
\caption{
\label{fig8} \footnotesize
(a):
Abundance ratios [V/Fe] versus [Ti/Fe] from ionized transitions of each
element.
The filled squares are from \cite{roe14} and the filled circle for
\hdeight\ is derived in this paper.
The horizontal and vertical (dotted) lines denote the solar abundance ratios
of each element.
The solid line represents a 45 degree slope.
(b):
Another V versus Ti correlation, this time from the neutral
species of each element.
(c): 
Means of [Ti I/Fe I] and [Ti II/Fe II] abundance
ratios plotted versus
means of [V I/Fe I] and [V II/Fe II] abundance ratios.
(d): 
Abundance ratios [Sc/Fe] versus [Ti/Fe] from
ionized transitions of each
element.
}                                                             
\end{figure}

It is also evident from Figure~\ref{fig7} 
that the abundances of the elements Sc, Ti and V are enhanced
with respect to solar values $-$ their abundance ratios exceed the 
[X/Fe]~=~0 horizontal dashed line in the figure. 
This result is not unique to \hdeight; it has been seen in other studies of
low metallicity stars (see e.g., \citealt{roe14, yon13}).  
We examine in Figure~\ref{fig8} the relationship between these abundances, 
first between the abundance ratio [\species{Ti}{ii}/\species{Fe}{ii}] and
[\species{V}{ii}/\species{Fe}{ii}] from the extensive abundance data 
of \cite{roe14}.
In panel (a) abundance ratios for stars with detectable \species{Ti}{ii} and 
\species{V}{ii} lines are shown along with the newly determined
abundance values [\species{Ti}{ii}/\species{Fe}{ii}] and 
[\species{V}{ii}/\species{Fe}{ii}] for \hdeight\ from this paper.\footnote{
Note that the new atomic data for \species{V}{i}, 
\species{V}{ii}, \species{Co}{i}, and \species{Ni}{i} that we applied to the 
spectrum of \hdeight\ were not available for the \cite{roe14} survey.}
Several trends are evident from this figure. 

First, the majority of the observed data are above solar values $-$ note the 
dashed lines indicating solar ratios. 
Caution is warranted here, and inclusion of upper limits would undoubtedly
add to the count of stars with sub-solar abundance ratios of these elements
with respect to Fe.
Second, while there is some abundance scatter it is clear that the 
relative abundances of Ti and V are correlated.
As a guide we have plotted a 45$^\circ$ (solid) line over the data. 
This line is consistent with the general trend of the \cite{roe14} data 
and intersects the abundance values for \hdeight. 
 
An additional examination of these two elements is made in 
panel (b) of Figure~\ref{fig8}, but now for the neutral species,
[\species{V}{i}/\species{Fe}{i}] and [\species{Ti}{i}/\species{Fe}{i}].
There are fewer observational data available $-$ nondetections and upper 
limits are not included $-$ but the trends are similar,
with most of the  
[\species{Ti}{i}/\species{Fe}{i}] ratios above solar 
and a strong correlation between the Ti and V abundance ratios. 
To get a sense of the observational limitations of the data and to further
examine the apparent correlation between these elements we show in 
panel (c) of Figure~\ref{fig8} averages of [\species{Ti}{i}/\species{Fe}{i}] 
and [\species{Ti}{ii}/\species{Fe}{ii}] versus the average of 
[\species{V}{i}/\species{Fe}{i}] and [\species{V}{ii}/\species{Fe}{ii}]. 
This averaging process included 117 stars with all abundance states 
measured. 
This compares to more than 200 stars with just ionized-species
abundances determined by \cite{roe14}.
This figure panel shows much less scatter than seen in the individual plots  
$-$ many of the ``outlier'' stars have vanished.  
Again, the average abundance ratios of almost all of Ti data are above 
solar, while a number of the V values show greater variation.
We again note  that the plot includes only detectable species and does
not include upper limits on V or Ti abundances in the data set of
\cite{roe14} which  includes a total of more than 300 halo stars. 
The data of this figure panel confirms and strengthens the abundance 
correlation between V and Ti in low metallicity stars. 

We noted earlier that the Sc abundance was also above solar in \hdeight. 
In panel (d) of Figure~\ref{fig8} we examine the abundance ratios of [Sc/Fe] 
versus [Ti/Fe], both determined from their ionized species.
While there is again scatter in the 
observational data, the same trends are seen in this plot as in the other
three plots of the figure.
Our new abundance ratios for [Sc/Fe] and [Ti/Fe] for \hdeight\ are also 
consistent with the majority trend of other metal-poor halo stars.  

In Figure~\ref{fig8} the \cite{roe14} data show two stars that consistently 
exhibit very low Sc, Ti, and V abundances:  BD+80~245 and CS~22169-035.
These stars have been analyzed by other authors, and BD+80~245 was the
first-discovered low metallicity star with substantially subsolar
$\alpha$-element abundances \citep{car97}.
All authors who have studied these stars reach very similar conclusions
on their Sc, Ti, and V abundances.
Here we average results from neutral and ionized species of an
element when both are available.
For BD+80~245, \cite{roe14} derived [Sc/Fe]~=~$-$0.40, [Ti/Fe]~=~$-$0.29 and
[V/Fe]~=~$-$0.36.
\cite{car97} obtained [Ti/Fe]~=~$-$0.26 and did not analyze Sc and V.
\cite{iva03} derived [Sc/Fe]~=~$-$0.42, [Ti/Fe]~=~$-$0.30 and
[V/Fe]~=~$-$0.39.
For CS~22169-035 \cite{roe14} found [Sc/Fe]~=~$-$0.40, [Ti/Fe]~=~$-$0.16 and
[V/Fe]~=~$-$0.44.
\cite{gir01} obtained somewhat larger abundance ratios,
[Sc/Fe]~=~$-$0.13 and [Ti/Fe]~=~$+$0.06.
However, two other studies are in accord with the \cite{roe14} values:
\cite{hon04} derived [Sc/Fe]~=~$-$0.33 and [Ti/Fe]~=~$-$0.13, while
\cite{cay04} derived [Sc/Fe]~=~$-$0.18 and [Ti/Fe]~=~$-$0.07.
We conclude that BD+80~245 and CS~22169-035 really are deficient in
Sc, Ti, and V with respect to Fe.
Their positions at the bottom of the V$-$Ti and Sc$-$Ti correlations
are confirmed in multiple studies.

The observational data illustrated in the figures
show that the three lightest Fe-group elements
(Sc, Ti and V) are correlated in metal-poor halo stars, and thus appear
to have closely related nucleosynthetic origins.
In the future it will be important to make further precise abundance
determinations in the metal-poor halo stars
utilizing the new atomic laboratory data for the iron-peak elements.

The production of Ti and V in core-collapse supernovae (CCSNe)
has been studied
in detail \citep[see e.g., discussion in][and references therein]{woo95},
and explosive stellar yields are available for different metallicities
\citep[e.g.,][]{woo95,lim00,rau02,heg10,pig13,chi13,nom13}.
In particular, $^{48}$Ti and $^{49}$Ti (which account for
73.7\% and 5.4\% of the solar Ti,
respectively) are mostly made in the explosive Si-burning stellar layers,
as $^{48}$Cr and $^{49}$Cr.
Additionally, $^{46}$Ti (8.2\% of the solar Ti) is predominantly 
produced by explosive O burning.
$^{47}$Ti is primarily formed  as itself by O-burning and as $^{47}$V in
explosive Si burning, but in general it is underproduced by
CCSNe compared to the previous Ti isotopes.
Finally, $^{50}$Ti is not efficiently made in baseline SN explosions.
$^{51}$V (99.75\% of the solar V) is produced  largely  as $^{51}$Mn in 
explosive Si-burning conditions and as $^{51}$Cr in explosive O burning.

As an example, in Figure~\ref{fig10} we show the production of Ti and V
in the ejecta of a 15M$_{\odot}$ (Z = 0.01) 
star CCSN \citep{pig13}.
The pre-supernova stellar structure has been calculated with the stellar
code GENEC \citep{egg08}.
The CCSN explosion simulations are based on the prescription by \cite{fry12}.
These results confirm the nucleosynthesis scenario described above for      
baseline one-dimensional CCSN models.                         
In particular, V is produced in explosive Si burning together 
with $^{48}$Ti and $^{49}$Ti, and in explosive O burning with $^{46}$Ti, with
similar V/Ti ratios for the two components of the SN ejecta.  
From Figure~\ref{fig10}, we also observe that the conditions yielding    
the most efficient production of $^{56}$Ni at mass coordinate 
M $\sim$ 2 M$_{\odot}$ is V- and Ti-poor.                     
Therefore, it is possible to have Fe production that is semi-detached from
V and Ti production in deep ejecta of baseline CCSN models, while synthesis
of V and Ti are more coupled in the same fusion conditions.          
Within this scenario, the elemental ratio V/Ti observed in metal poor       
stars can be used to explore the properties of explosive Si-burning
and O-burning regions.                                        
The observed correlation of V and Ti may be a signature of this 
common production.                                            
This makes clear how important it is to have consistent       
observations with small errors and reliable atomic physics employed for     
a large sample of stars.                                      
In principle, the theoretical elemental ratios from stellar   
models could be compared with observations to directly constrain 
specific regions of the                                       
CCSN ejecta where Ti and V are produced.                      
The real stellar abundance  scatter, above any  observational 
errors, would also provide crucial information about          
different nucleosynthetic components                          
hiding in these old stars.                                    
This work started with HD 84937 is an important step in this  
direction that needs to be extended to other stars in the future.
                                                              
In panel (d) of Figure~\ref{fig8} we have seen that Sc production 
seems also to be correlated with Ti in the observational data.              
Such a correlation is not obvious from the theoretical point of view.
Sc can be synthesized in the same astrophysical site as Ti, but it has a   
more complex production history.                                   
The mono-isotopic Sc can be made as radiogenic $^{45}$Ti in explosive 
O-burning and Si-burning layers together with Ti and V \citep{woo95}.   
However, Sc can be also produced more efficiently as a result of neutrino 
feedback (\eg, \citeauthor{woo95}, \citealt{fro06a}), 
or in $\alpha$-rich freeze-out conditions. 
We will return to a discussion of Sc in \S\ref{gce}.
Nevertheless, it is clear that with all these different possible 
components the ratios Sc/Ti and Sc/V potentially are a less direct 
constraint than the Ti/V ratio.
Interestingly, despite all these different nucleosynthesis paths to make Sc,
theoretical stellar yields of CCSNe severely underestimate the observed Sc  
abundances compared to Fe in the early Galaxy (see also the next section).

\begin{figure}
\epsscale{0.8}
\plotone{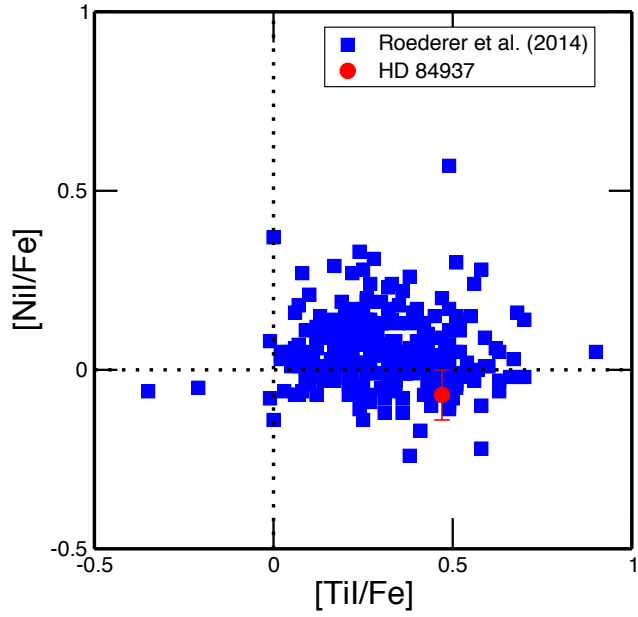}
\caption{
\label{fig9} \footnotesize
Abundance ratios [Ni/Fe] versus [Ti/Fe] from neutral transitions of each
element.
The symbols and lines are as in Figure~\ref{fig8}.
}
\end{figure}

\begin{figure}
\centering
\resizebox{12cm}{!}{\rotatebox{0}{\includegraphics{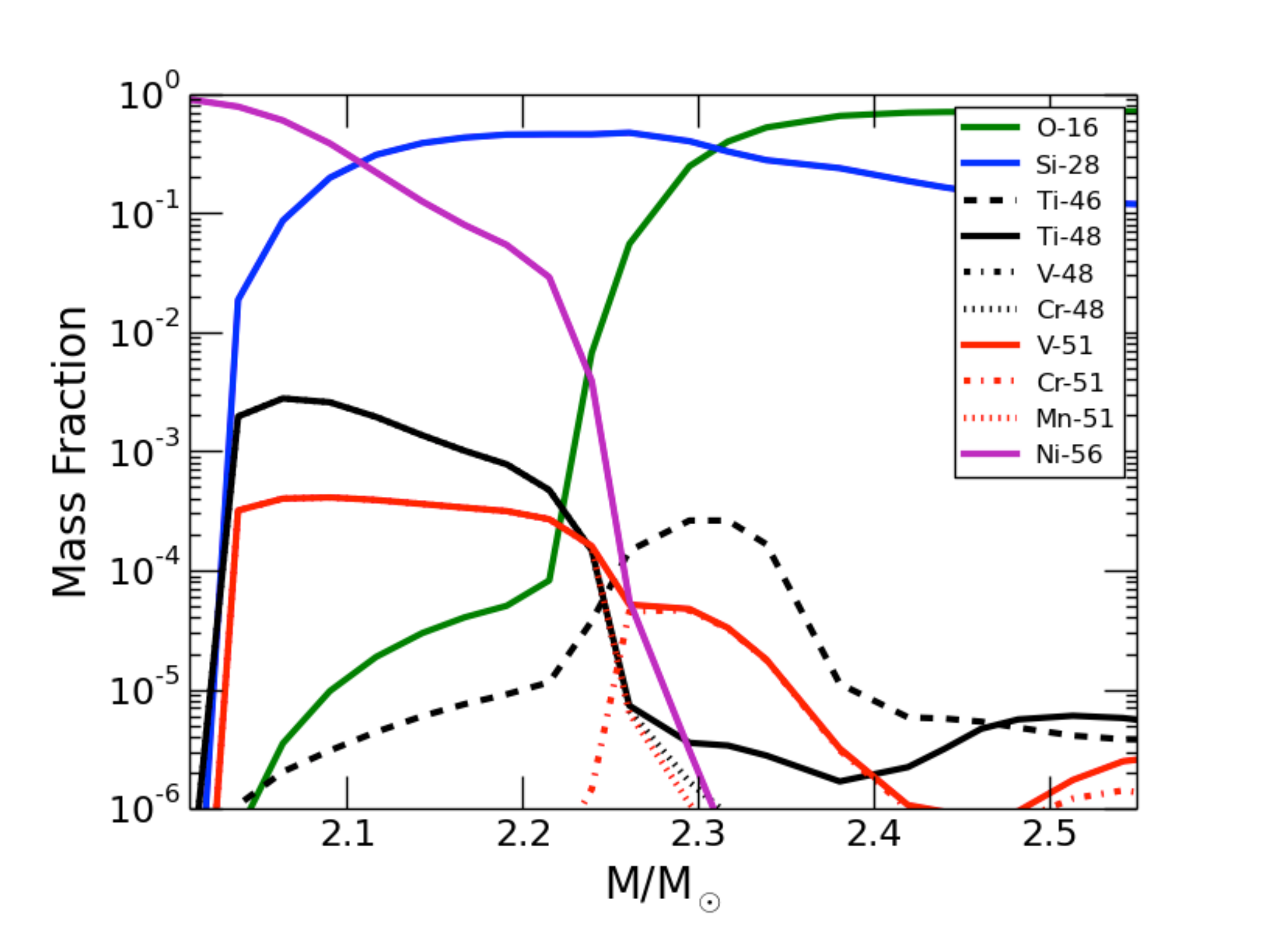}}}
\resizebox{12cm}{!}{\rotatebox{0}{\includegraphics{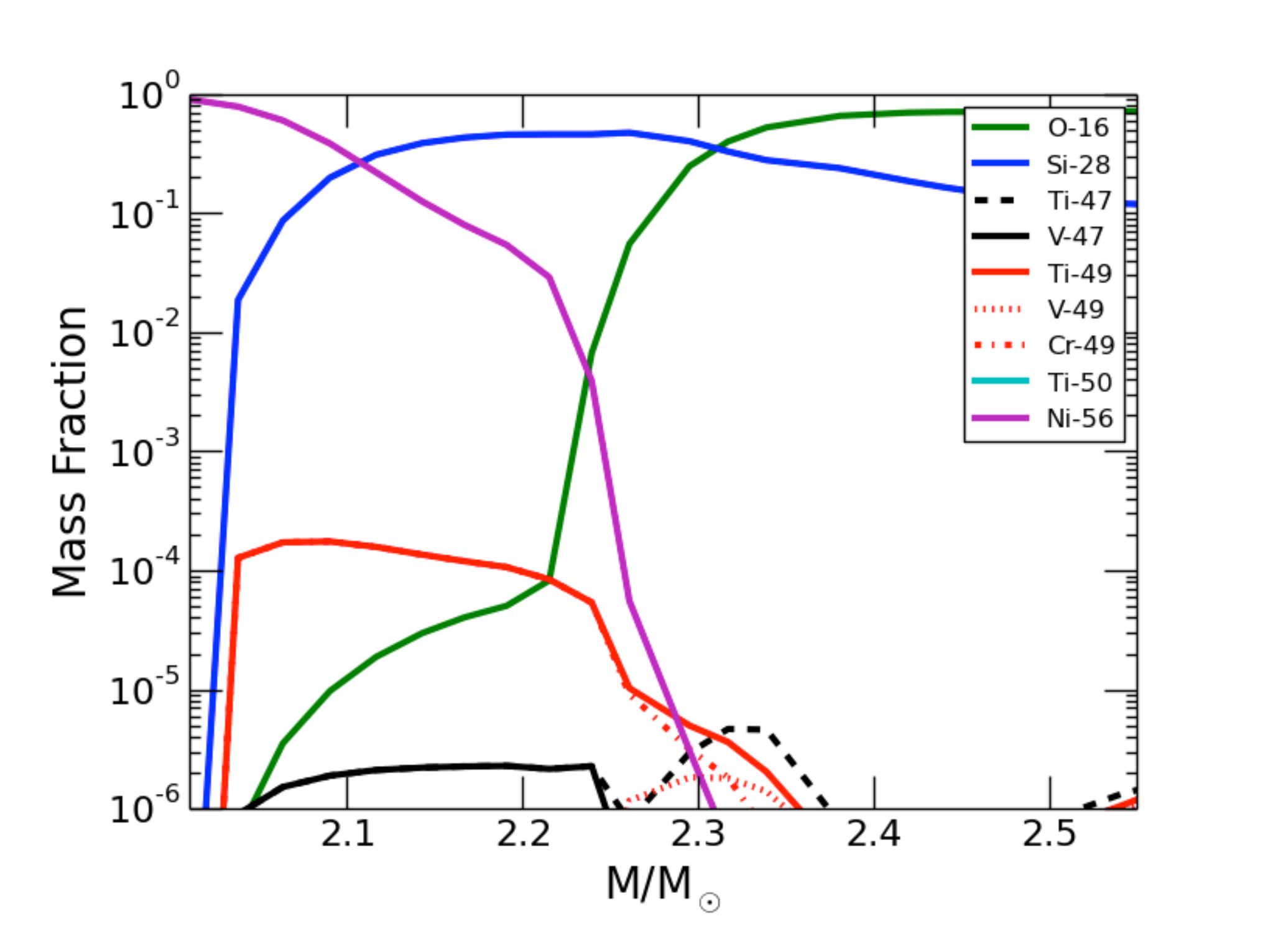}}}
\caption{
\label{fig10} \footnotesize
Upper panel: Isotopic abundances in the explosive Si-burning and O-burning
ejecta of the $15\mathrm{M_{\odot}}$ SN model \citep[][]{pig13}.
Shown are profiles for $^{16}$O, $^{28}$Si, $^{46}$Ti, $^{48}$Ti and
its radiogenic parent isotopes $^{48}$V and $^{48}$Cr, $^{51}$V and
its radiogenic parent isotopes $^{51}$Cr and $^{51}$Mn, and $^{56}$Ni.
The unstable isotope $^{56}$Ni will decay to $^{56}$Co and finally
to $^{56}$Fe, which is most of the Fe SN ejecta.
Lower panel: The same of the upper panel, but for $^{47}$Ti and its
radiogenic parent isotope $^{47}$V, $^{49}$Ti and its radiogenic
parent isotopes $^{49}$V and $^{49}$Cr, and for $^{50}$Ti.
}
\end{figure}

In addition to V and Sc we also examined other possible
correlations of iron-peak elemental abundances with those of Ti.
Specifically, we employed the \species{Ni}{i} abundance data from
\cite{roe14} and in Figure~\ref{fig9}  we have plotted their [Ni I/Fe] ratios
as a function of [Ti I/H].
Inspection of this figure reveals a scatter plot with no obvious correlations.
The [Ni I/Fe] value of \hdeight\ is unremarkable in this figure. 
We also note the two very low abundance Ti stars (BD+80 245 and CS 22169-035)
at the far left of Figure~\ref{fig9}.  As noted above, these stars 
are deficient in Ti, Sc and V with respect to iron; but their Ni abundances
do not show a similar low abundance 
trend and are consistent with that of \hdeight.\ 
 This lends further support to the idea that
Ni is synthesized  in a different process or environment than the iron
peak elements Sc-Ti-V.
We made a similar analysis (with Ti) 
for the element Mn and found a similar result $-$
no correlation between abundances derived from \species{Mn}{i}
and \species{Ti}{i}.
Thus, among the iron-peak elements only Sc and V appear
to be correlated with Ti production.

As a further comparison we examined the $\alpha$-group 
element Mg in the \cite{roe14} data. 
The $\alpha$ elements are thought to be produced mainly in massive stars. 
Sneden et al. (2008) compiled [Mg/Fe] data published in surveys to that
date, showing that this abundance ratio shows very little star-to-star 
scatter at all metallicities, even for [Fe/H]~$<$~$-$3.
and that at early times in the Galaxy, both Fe and Mg were being 
synthesized in the same types of stars.
An analysis of Mg I with Ti I from the \cite{roe14} shows no such 
behavior and only a scatter plot.

\section{IMPLICATIONS FOR EARLY GALACTIC NUCLEOSYNTHESIS\label{gce}}

For solar-neighborhood stars with metallicities of [Fe/H]~$\lesssim$~$-1$,
the Fe-group elements were predominantly made by CCSNe.
In these early stages, thermonuclear supernovae (Type Ia Supernovae, 
SNe Ia; \citealt{hil13} and references therein) did not have time to 
efficiently contribute to the galactic chemical inventory \citep{mat86,kob09}.
Therefore, by comparing the observations of Fe-group elements in \hdeight\ 
and other metal poor stars with Galactic chemical evolution (GCE) models,
it is possible to provide direct constraints on CCSN predictions.
In particular, these elements are ideal sources of information 
about the details of SNe explosions.                        
While light elements like O and Mg are mostly made in the pre-SN stage,
Fe-group elements are mostly made during the explosion, in the deepest part 
of the ejecta \citep[e.g.,][for a review]{nom13}.                           
Although there has been great progress in multi-dimensional simulations of 
SN explosions in recent years, the CCSN engine 
\citep[][and references therein]{hix14} and subsequent propagation 
of the SN shock through the progenitor structure \citep{won15} 
are uncertain at present.
Modern CCSN yield sets are still calculated with one-dimensional codes
(\eg, \citealt{nom13}), and multi-dimensional effects have not been 
included in the GCE models yet.

During a SN explosion, the elements that are synthesized in different
stellar layers should mix to some extent, and some fraction of this mixed
material will fall back onto the remnant (a neutron star or a black hole).
Recent nucleosynthesis yields have been calculated with this fallback (\ie,
mass cut; \citealt{kob06}, \citealt{heg10}, \citealt{lim12}) and 
mixing (\citeauthor{kob06}).
In this last paper's yields, the ejected iron masses are constrained
with independent parameters, \ie, the observed light curves and
spectra of nearby supernovae \citep{nom13}.
Hypernovae (with stellar masses of 
$M \ge 25M_\odot$) have explosion energies ten times higher than 
normal SNe. 
In our GCE models, we assume a 50\% hypernovae fraction, 
as did \citeauthor{kob06}

In core-collapse supernova ejecta, Cr and Mn are synthesized in the
incomplete-Si burning regions, while Co, Ni, Cu, and Zn are synthesized
in the more central, complete-Si burning regions.
Therefore, [(Cr,Mn)/Fe] and [(Co,Ni,Cu,Zn)/Fe] ratios depend on the
mixing-fallback parameters.
On the other hand, the ratios among them, e.g., [Mn/Cr], are more
sensitive to the number of electrons per nucleon, $Y_{\rm e}$
(see \S3.3 of \citealt{nom13} for the details).
Despite these complicated parameter dependencies, the GCE models are in good 
agreement with observations except for Sc, Ti, and V \citep{kob06,kob11b}.

In GCE models, the contributions from various supernovae 
with different masses, metallicites, and energies are integrated according 
to the star formation history of the system considered. 
In this paper, we use 3 GCE models for the solar neighborhood: \citet{kob06}, 
\citet{kob11b}, and a new model including a jet effect to the 2011 model. 
In reality, hypernovae should be jet-like explosions, and \citet{mae03} 
showed that some Fe-peak elements are significantly enhanced by the 
the strong $\alpha$-rich freeze-out due to high temperatures and 
high entropies in complete-Si burning with bipolar models.
However, nucleosynthesis yields with such 2D models have been calculated only
for a small number of parameter sets.
Therefore, we estimate a typical enhancement from the yields in 
\citet[][$40M_\odot$, $Z=0$, model C]{tom09} and apply constant factors,
$+1.0, 0.45, 0.3, 0.2$, and $0.2$~dex for [(Sc, Ti, V, Co, and Zn)/Fe],
respectively, to the hypernova yields with other parameters.
Similar effects are expected also for 3D models that naturally show a
large inhomogeneity \citep[e.g.,][]{jan12,bru13,bur13}, although the
nucleosynthesis yields are not available yet.

All GCE models assume an infall of primordial gas from 
outside the disk region, star formation proportional to the gas fraction, 
and no outflow.
The model parameters are determined to meet the observed metallicity 
distribution function.
The \citet{sal55} IMF and \citet{kro01} IMF are adopted for the 2006 and 
2011/2015 models, respectively.
Note that the contributions from AGB stars are also included in the  
2011/2015 models, but the elemental abundances of Fe-peak elements are 
not affected at all.
In all models, the contributions from SNe Ia are included as
Chandrasekhar-mass explosions from single  degenerate systems
(\citealt{kob98,kob09}, see e.g., \citealt{kob15} for other progenitor models).
The delay-time distributions of SNe Ia are slightly different between the
2006 and 2011 models.
The nucleosynthesis yields of a deflagration are adopted in the 
2006/2011 models, 
while those of a delayed
detonation \citep[the N100 model]{sei13} are adopted in 
the 2015 model.

\begin{figure}
\epsscale{1.1}
\plotone{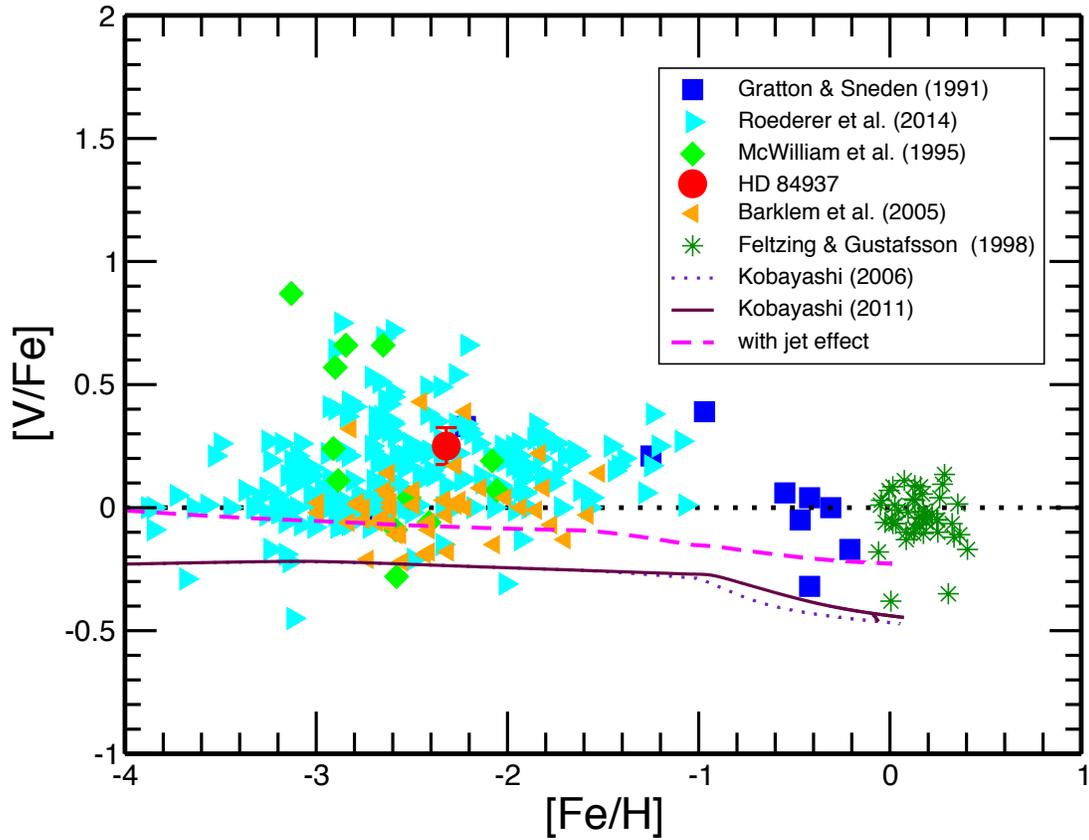}
\caption{
\label{fig11} \footnotesize
Abundance ratios [V/Fe] plotted as function of [Fe/H] metallicity
from \cite{gra91}, \cite{roe14}, \cite{mcw95}, \cite{bar05}, \cite{fel98}. 
The symbols are defined in the figure legend. The solid circle is the 
abundance value for HD 84937 derived in this paper.  
Overlaid on the figure are GCE models from Kobayashi (2006; 2011; and this
paper). See text for detailed discussion.
}
\end{figure}

To illustrate these GCE trends of the iron-peak elements, particularly      
at low metallicities, we first focus in detail on the [V/Fe] abundance 
ratios as a function of [Fe/H] in Figure~\ref{fig11}. 
The observational data span decades and include large-sample surveys such
as \cite{gra91,mcw95,fel98,bar05,roe14}.                         
We again note that these earlier observations did not have access to the 
most up-to-date laboratory data and had to rely on then current literature 
values. 
Nevertheless, the overall trends are clear.                                     
First, most of the [V/Fe] abundance values in these low metallicity stars 
are above the solar value $-$ note the dashed horizontal line in 
Figure~\ref{fig11}.                                         
The data also show a rise at low metallicities; our new value of [V/Fe]  
for \hdeight\ is consistent with that trend.                   
We also note the large scatter in abundance values for these stars at 
lower metallicities.
Superimposed on these data are several GCE model curves.      
These include \cite{kob06,kob11b} and the new model with the jet effects, 
indicated by the 
dashed line in Figure~\ref{fig11}.                                        
It is seen in the figure that including additional physics effects has led to 
increasing agreement between theory and observations.

\begin{figure}
\epsscale{1.0}
\plotone{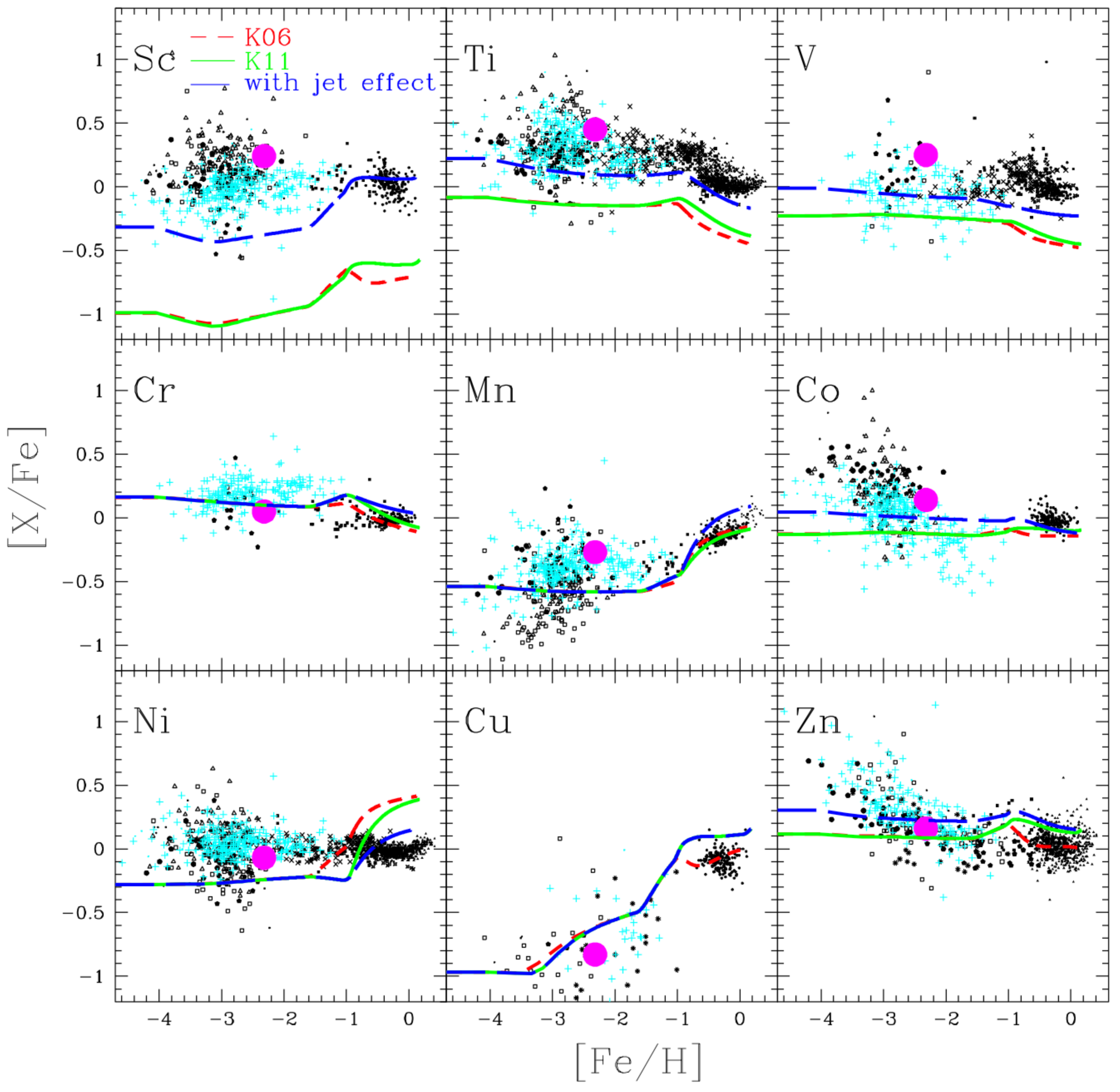}
\caption{
\label{fig12} \footnotesize
Abundance ratios of iron peak elements plotted as functions of [Fe/H] 
metallicity from a number of observational studies (see text for discussion).
The solid magenta circles represent the \hdeight\ abundance ratios 
derived in this paper.
Overlaid on the figure are GCE models from \cite{kob06,kob11b} and this
paper. 
See text for detailed discussion.
}
\end{figure}

Figure~\ref{fig12} gives an overview of the evolution of all Fe-peak elements
in the solar neighborhood.
The large filled circles are the mean abundances of \hdeight\ from
Table~\ref{tab5}.
The black dots represent observational data before 2014
(\citealt{ful00,pri00,gra03,red03,cay04,hon04,fel07,nis07,sai09,coh13,yon13})
and the cyan plus signs are from \cite{roe14}.
Elements Sc, Ti, V, and Cr have some abundance information from both neutral
and ionized species in various literature sources.
In Figure~\ref{fig12} we have chosen to plot only abundances derived from
\species{Sc}{ii}, \species{Ti}{i}, \species{V}{i}, and \species{Cr}{ii}.
For Ti, V, Cr, Ni, Cu, and Zn there is excellent accord
between the survey results of \cite{roe14} and those of the other included
literature studies, but systematic offsets of $\sim$0.2$-$0.3~dex
are apparent for Sc, Mn and Co.
The reasons for these shifts deserve systematic investigation in future studies.

The difference in the adopted IMFs is negligible, and the 
2006 (red short-dashed lines) and 2011 (green solid lines) models give 
almost the same trends at [Fe/H] $\lesssim -1$.  
The small difference at [Fe/H] $\gtrsim -1$ between the 2006 and 2011 
models is caused by the small difference in the delay-time distributions 
of SNe Ia.
The blue dashed lines of Figure~\ref{fig12} include the effects of jet-like
explosions of hypernovae, and one sees that these models are much closer
to the observational data especially for Sc, Ti, V, and perhaps Zn.
In addition, the $\nu$-process also enhances Sc, V, and Ti
isotopes \iso{46,47,49,50}{Ti} in the yields calculated 
in \citet{kob11a}.
We note that currently the new GCE model illustrated in Figure~\ref{fig11}
and Figure~\ref{fig12} underproduces the observed Sc, V, and Ti abundance
ratios in \hdeight\ and in the bulk of stars.
Nevertheless, the interplay between the multidimensional effects and
additional neutrino effects (e.g., the $\nu$-processes), may increase the
predicted production of Sc, Ti, and V abundances making them closer
to those that are observed.

Detailed explorations of models and abundances for 
all Fe-group elements are beyond the scope of this paper, but
a few other comments here are appropriate.
Cr abundances are consistent with all these models, which has been shown in
the comparison with the \species{Cr}{ii} observations in \citet{kob06}.
[Mn/Fe] seems $\sim 0.1-0.2$ dex lower in the models than 
in the observations at [Fe/H] $\lesssim -1$, which can easily solved with 
varying $Y_{\rm e}$.
Both in the observations and GCE models, the [Mn/Fe] ratio shows an
increasing trend from [Fe/H]~$\sim -1$ to $\sim 0$ because Mn is produced
more by SNe Ia than by core-collapse supernovae.
This Mn increase cannot be reproduced with other SN Ia models such as
double degenerate systems or double detonations \citep{sei13b,kob15}.
Ni/Fe] seems $\sim 0.3$ dex lower in the models than in 
the observations at [Fe/H] $\lesssim -1$, which may be difficult to solve 
because Ni and Zn have an opposite response to $Y_{\rm e}$.
Ni is overproduced by SNe Ia in the short-dashed and solid lines at
[Fe/H]~$\gtrsim$~$-1$.
As noted in \citet{kob06}, with delayed detonations, this Ni
overproduction problem can be eased, 
which is included in the 2015 model (long-dashed lines).
At [Fe/H] $\lesssim$ -2.5, the observed [Co/Fe] and [Zn/Fe] ratios
show a rising trend at lower metallicities.
The observed Cu abundances are roughly consistent with the models.
As demonstrated by \citet{ume02} and \citet{kob06}, higher explosion energies
lead to higher values of
[(Co,Zn)/Fe], so that an even higher hypernovae fraction may be 
required at low metallicities.

\section{SUMMARY AND CONCLUSIONS\label{sum}}

We have conducted an extensive Fe-group elemental abundance analysis
in the metal-poor main-sequence turnoff star \hdeight, involving 
nearly 1200 lines of 17 species.
Derived abundances of neutrals and ions are the same within mutual 
uncertainties for all seven Fe-group elements with two
species available for study.
This strongly suggests that standard Saha ionization balance holds
for \hdeight, and that the observed abundance ratios among Fe-group
elements can be treated as true measures of prior nucleosynthesis events.

The $\sim$0.3~dex overabundances of Sc, Ti and V relative 
to Fe in \hdeight\ are of greatest nucleosynthetic interest.
An examination of the abundances of Sc, Ti and V suggest a correlation
of their abundance enrichments in the early Galaxy.
This may be due to a common nucleosynthesis origin for Ti and V in the 
CCSN ejecta, while the production of Sc is more complex. 
A more detailed analysis to clearly identify what components are 
dominating the production of Sc is needed.

GCE models cannot yet reproduce all of the observed abundance trends as a 
function of [Fe/H] for these iron-peak elements, particularly 
at low metallicities. 
We explore the impact of new stellar models, and we discuss how 
observations can be used to constrain theoretical models of CCSNe 
and hypernovae. 
For \hdeight\ we have derived [(Sc,Ti,V/Fe)]~$\simeq$~$+$0.3, well
above the observed uncertainties of $\lesssim$0.1 dex from the species
$\sigma$ values.
In the future, additional precise abundance determinations in 
metal-poor halo stars, taking advantage of new atomic laboratory data for 
the Fe-peak elements, should be undertaken. 
They  suggest that with the additional inclusion of physics effects,  the 
observational and theoretical agreement will improve.

\acknowledgments

We thank Tim Beers, John Norris, Craig Wheeler, Ian Roederer 
and David Yong for helpful discussions about this work.
Our referee and Pat Scott made very useful comments on the 
submitted manuscript, which led to improvement in the paper.
This work has been supported in part by NASA grant NNX10AN93G (J.E.L.), 
by NSF grant AST-1211055 (J.E.L.) and NSF grant AST-1211585 (C.S.).  
JJC was supported in part by the   
JINA Center for the Evolution of the Elements, 
supported by the National Science Foundation under Grant No. PHY-1430152.  
MP acknowledges significant support to NuGrid from NSF grants PHY 09-22648 
(Joint Institute for Nuclear Astrophysics, JINA), NSF grant PHY-1430152 
(JINA Center for the Evolution of the Elements) and EU MIRGCT-2006-046520. 
MP acknowledges the support from the ``Lendlet-2014'' Programme of the 
Hungarian Academy of Sciences (Hungary) and from SNF (Switzerland).



\clearpage

\clearpage

\clearpage

\clearpage

\clearpage

\clearpage

\clearpage

\clearpage

\clearpage

\clearpage
\begin{center}
\begin{deluxetable}{lccrccr}
\tabletypesize{\footnotesize}
\tablewidth{0pt}
\tablecaption{Species Abundances\label{tab1}}
\tablecolumns{7}
\tablehead{
\colhead{Element}             &
\colhead{log $\epsilon$}      &
\colhead{$\sigma$}            &
\colhead{num}                 &
\colhead{log $\epsilon$}      &
\colhead{$\sigma$}            &
\colhead{num}                 \\
\colhead{}                    &
\colhead{\sc i}               &
\colhead{\sc i}               &
\colhead{\sc i}               &
\colhead{\sc ii}              &
\colhead{\sc ii}              &
\colhead{\sc ii}              
}
\startdata
\multicolumn{7}{c}{solar photosphere} \\
Sc  & \nodata              & \nodata & \nodata &   3.167 $\pm$ 0.014 &   0.056 &      15 \\
Ti  &   4.973 $\pm$ 0.003  &   0.040 &     168 &   4.979 $\pm$ 0.005 &   0.054 &      43 \\
V   &   3.956 $\pm$ 0.004  &   0.037 &      93 &   3.950 $\pm$ 0.010 &   0.050 &      15 \\
Cr  &   5.650 $\pm$ 0.007  &   0.068 &      87 &   5.630 $\pm$ 0.030 &   0.153 &      13 \\
Mn  &   5.451 $\pm$ 0.011  &   0.050 &      19 &   5.400 $\pm$ 0.000 &   0.000 &       1 \\
Fe  &   7.522 $\pm$ 0.004  &   0.047 &     152 &   7.511 $\pm$ 0.021 &   0.082 &      16 \\
Co  &   4.955 $\pm$ 0.007  &   0.059 &      82 & \nodata             & \nodata & \nodata \\
Ni  &   6.277 $\pm$ 0.004  &   0.055 &      76 & \nodata             & \nodata & \nodata \\
Cu  &   4.210 $\pm$ 0.013  &   0.033 &       7 & \nodata             & \nodata & \nodata \\
Zn  &   4.610 $\pm$ 0.050  &   0.087 &       3 & \nodata             & \nodata & \nodata \\

\multicolumn{7}{c}{HD 84937} \\
Sc  & \nodata              & \nodata & \nodata &   1.081 $\pm$ 0.007 &   0.035 &      25 \\
Ti  &   3.122 $\pm$ 0.007  &   0.054 &      54 &   3.081 $\pm$ 0.007 &   0.087 &     147 \\
V   &   1.890 $\pm$ 0.030  &   0.070 &       9 &   1.871 $\pm$ 0.009 &   0.075 &      68 \\
Cr  &   3.304 $\pm$ 0.018  &   0.098 &      31 &   3.437 $\pm$ 0.008 &   0.097 &      78 \\
Mn  &   2.843 $\pm$ 0.019\tablenotemark{a}  &   
                               0.083 &      19 &   2.881 $\pm$ 0.008 &   0.026 &      10 \\
Fe  &   5.202 $\pm$ 0.003  &   0.069 &     446 &   5.193 $\pm$ 0.006 &   0.059 &     105 \\
Co  &   2.787 $\pm$ 0.008  &   0.068 &      66 &   2.771 $\pm$ 0.020 &   0.079 &      15 \\
Ni  &   3.888 $\pm$ 0.008  &   0.068 &      77 &   3.890 $\pm$ 0.040 &   0.100 &       8 \\
Cu  &   1.060 $\pm$ (0.03)\tablenotemark{b} &  (0.07)\tablenotemark{b} 
                                     &       2 & \nodata             & \nodata & \nodata \\
Zn  &   2.450 $\pm$ 0.037  &   0.075 &       4 & \nodata             & \nodata & \nodata \\
\enddata

\tablenotetext{a}{The mean abundance from \species{Mn}{i} lines has been computed
                  with exclusion of the 4030, 4033, 4034~\AA\ resonance triplet.}

\tablenotetext{b}{The Cu $\sigma$ value has been assigned to more realistically
                  represent its abundance uncertainty in \hdeight, and the
                  mean abundance error follows from that and the number of 
                  lines.}

\end{deluxetable}

\end{center}

\clearpage
\begin{center}
\begin{deluxetable}{lccccccc}
\tabletypesize{\footnotesize}
\tablewidth{0pt}
\tablecaption{Solar log $\epsilon$ Abundance Comparisons\label{tab2}}
\tablecolumns{8}
\tablehead{
\colhead{Element}                           &
\colhead{new\tablenotemark{a}}              &
\colhead{Sco15\tablenotemark{b}}            &
\colhead{Sco15\tablenotemark{c}}            &
\colhead{Sco15\tablenotemark{d}}            &
\colhead{Sco15\tablenotemark{e}}            &
\colhead{Asp09\tablenotemark{f}}            &
\colhead{Lod09\tablenotemark{g}}            \\
\colhead{}                                  &
\colhead{}                                  &
\colhead{HMLTE}                             &
\colhead{HMNLTE}                            &
\colhead{3D}                                &
\colhead{rec}                               &
\colhead{}                                  &
\colhead{}     
}
\startdata
\multicolumn{8}{c}{neutral species} \\
Sc  & \nodata &    3.12 &    3.28 &    3.21 &    3.16 &    3.15 &    3.07 \\
Ti  &    4.97 &    4.96 &    4.99 &    4.94 &    4.93 &    4.95 &    4.93 \\
V   &    3.96 &    3.97 &    4.07 &    3.98 &    3.89 &    3.93 &    3.98 \\
Cr  &    5.65 &    5.65 &    5.66 &    5.62 &    5.62 &    5.64 &    5.66 \\
Mn  &    5.45 &    5.43 &    5.47 &    5.43 &    5.42 &    5.43 &    5.50 \\
Fe  &    7.52 &    7.51 &    7.52 &    7.46 &    7.47 &    7.50 &    7.47 \\
Co  &    4.96 &    4.94 &    4.99 &    4.96 &    4.93 &    4.99 &    4.89 \\
Ni  &    6.28 &    6.24 &    6.24 &    6.20 &    6.20 &    6.22 &    6.22 \\
Cu  &    4.21 &    4.22 &    4.21 &    4.16 &    4.18 &    4.19 &    4.27 \\
Zn  &    4.61 &    4.55 &    4.53 &    4.52 &    4.56 &    4.56 &    4.65 \\

\multicolumn{8}{c}{ionized species} \\
Sc  &    3.19 &    3.20 &    3.19 &    3.17 &    3.16 &    3.15 &    3.07 \\
Ti  &    4.97 &    4.97 &    4.97 &    4.97 &    4.93 &    4.95 &    4.93 \\
V   & \nodata &    4.03 &    4.03 & \nodata &    3.89 &    3.93 &    3.98 \\
Cr  &    5.63 &    5.63 &    5.63 &    5.65 &    5.62 &    5.64 &    5.66 \\
Mn  & \nodata & \nodata & \nodata & \nodata &    5.42 &    5.43 &    5.50 \\
Fe  &    7.46 &    7.46 &    7.46 &    7.51 &    7.47 &    7.50 &    7.47 \\
Co  & \nodata & \nodata & \nodata & \nodata &    4.93 &    4.99 &    4.89 \\
Ni  & \nodata &    6.24 &    6.24 &    6.23 &    6.20 &    6.22 &    6.22 \\
Cu  & \nodata & \nodata & \nodata & \nodata &    4.18 &    4.19 &    4.27 \\
Zn  & \nodata & \nodata & \nodata & \nodata &    4.56 &    4.56 &    4.65 \\
\enddata

\tablenotetext{a}{new = this study}
\tablenotetext{b}{Sco15 = \cite{sco15} for Sc$-$Ni, \cite{gre15} for Cu, Zn;
                  HMLTE = Holweger \& M{\" u}ller (1974) model, LTE; these 
                  values are from P. Scott (2015, private communication)}
\tablenotetext{c}{Sco15, HMNLTE = Holweger \& M{\" u}ller model, NLTE 
                  corrections}
\tablenotetext{d}{Sco15, 3D = 3-dimensional model atmosphere}
\tablenotetext{e}{Sco15, rec = final recommended photospheric value from 
                  both species}
\tablenotetext{f}{Asp09 = recommended photospheric value from \cite{asp09}}
\tablenotetext{g}{$\rm L$od09 = recommended meteoritic value from \cite{lod09}}

\end{deluxetable}

\end{center}

\clearpage
\begin{center}
\begin{deluxetable}{lcccccc}
\tabletypesize{\footnotesize}
\tablewidth{0pt}
\tablecaption{Line Abundances for \ion{Fe}{1} and \ion{Fe}{2}\label{tab3}} 
\tablecolumns{7}
\tablehead{
\colhead{$\lambda$}           &
\colhead{species}             &
\colhead{$\chi$}              &
\colhead{log $gf$}            &
\colhead{source\tablenotemark{a}} &
\colhead{log $\epsilon$}      &
\colhead{log $\epsilon$}      \\
\colhead{\AA}                 &
\colhead{}                    &
\colhead{eV}                  &
\colhead{}                    &
\colhead{}                    &
\colhead{Sun}                 &
\colhead{HD 84937}            
}
\startdata
2297.787 & FeI & 0.052 & $-$1.10 & Obr91 & \nodata & 5.27 \\ 
2298.658 & FeI & 0.110 & $-$2.42 & Obr91 & \nodata & 5.17 \\
2299.220 & FeI & 0.087 & $-$1.55 & Obr91 & \nodata & 5.27 \\
2308.999 & FeI & 0.110 & $-$1.39 & Obr91 & \nodata & 5.37 \\
2369.455 & FeI & 0.110 & $-$2.19 & Obr91 & \nodata & 5.12 \\
2371.430 & FeI & 0.087 & $-$1.95 & Obr91 & \nodata & 5.37 \\
2374.518 & FeI & 0.121 & $-$2.10 & Obr91 & \nodata & 5.29 \\
2389.970 & FeI & 0.087 & $-$1.57 & Obr91 & \nodata & 5.24 \\
2438.182 & FeI & 0.858 & $-$1.25 & Obr91 & \nodata & 5.12 \\
2453.476 & FeI & 0.914 & $-$0.92 & Obr91 & \nodata & 5.17 \\
\enddata

\vspace*{0.2in}
Note –this Table is available in its entirety via the link to the 
machine-readable version online.

\tablenotetext{a}{Den14 = \cite{den14};
                  Obr91 = \cite{obr91};
                  Ruf14 = \cite{ruf14};
                  NIST = NIST database}

\end{deluxetable}

\end{center}

\clearpage
\begin{center}
\begin{deluxetable}{lccrccc}
\tabletypesize{\footnotesize}
\tablewidth{0pt}
\tablecaption{Line Abundances for Other Fe-Group Species\label{tab4}} 
\tablecolumns{7}
\tablehead{
\colhead{$\lambda$}           &
\colhead{species}             &
\colhead{$\chi$}              &
\colhead{log $gf$}            &
\colhead{source\tablenotemark{a}} &
\colhead{log $\epsilon$}      &
\colhead{log $\epsilon$}      \\
\colhead{\AA}                 &
\colhead{}                    &
\colhead{eV}                  &
\colhead{}                    &
\colhead{}                    &
\colhead{Sun}                 &
\colhead{HD 84937}            
}
\startdata
2552.354 & ScII & 0.022 &    0.033 &  Law89 & \nodata & 1.00 \\
2563.190 & ScII & 0.000 & $-$0.575 &  Law89 & \nodata & 1.10 \\
3353.717 & ScII & 0.315 &    0.251 &  Law89 & \nodata & 1.02 \\
3359.675 & ScII & 0.008 & $-$0.738 &  Law89 & \nodata & 1.10 \\
3368.932 & ScII & 0.008 & $-$0.373 &  Law89 & \nodata & 1.10 \\
3535.714 & ScII & 0.315 & $-$0.465 &  Law89 & \nodata & 1.05 \\
3567.693 & ScII & 0.000 & $-$0.476 &  Law89 & \nodata & 1.07 \\
3576.336 & ScII & 0.008 &    0.007 &  Law89 & \nodata & 1.12 \\
3580.920 & ScII & 0.000 & $-$0.149 &  Law89 & \nodata & 1.09 \\
3589.628 & ScII & 0.008 & $-$0.574 &  Law89 & \nodata & 1.09
\enddata

\vspace*{0.2in}
Note –this Table is available in its entirety via the link to the 
machine-readable version online.

\tablenotetext{a}{$\rm L$aw89 = \cite{law89};
                  Sob07 = \cite{sob07};
                  Gur10 = \cite{gur10};
                  Nil06 = \cite{nil06};
                  Den11 = \cite{den11};
                  NIST = NIST database;
                  Roe14 = \cite{roe14}
}

\end{deluxetable}

\end{center}

\clearpage
\begin{center}
\begin{deluxetable}{lrrrrrrr}
\tabletypesize{\footnotesize}
\tablewidth{0pt}
\tablecaption{Final Relative Abundances\label{tab5}}
\tablecolumns{8}
\tablehead{
\colhead{El}                  &
\colhead{Z}                   &
\colhead{[X/H]}               &
\colhead{$\sigma$}            &
\colhead{[X/H]}               &
\colhead{$\sigma$}            &
\colhead{[X/H]}               &
\colhead{[X/Fe]}              \\
\colhead{}                    &
\colhead{}                    &
\colhead{\sc i}               &
\colhead{\sc i}               &
\colhead{\sc ii}               &
\colhead{\sc ii}               &
\colhead{mean}                &
\colhead{mean}                
}
\startdata
Sc & 21 & \nodata & \nodata & $-$2.09 &    0.04 & $-$2.09 & $+$0.23 \\
Ti & 22 & $-$1.85 &    0.05 & $-$1.90 &    0.09 & $-$1.87 & $+$0.45 \\
V  & 23 & $-$2.07 &    0.07 & $-$2.08 &    0.08 & $-$2.07 & $+$0.25 \\
Cr & 24 & $-$2.35 &    0.10 & $-$2.19 &    0.10 & $-$2.27 & $+$0.05 \\
Mn & 25 & $-$2.61 &    0.08 & $-$2.57\tablenotemark{a}
                                      &    0.03 & $-$2.59 & $-$0.27 \\
Fe & 26 & $-$2.32 &    0.07 & $-$2.32 &    0.06 & $-$2.32 &    0.00 \\
Co & 27 & $-$2.17 &    0.07 & $-$2.18\tablenotemark{a}
                                      &    0.08 & $-$2.18 & $+$0.14 \\
Ni & 28 & $-$2.39 &    0.07 & $-$2.39\tablenotemark{a}
                                      &    0.10 & $-$2.39 & $-$0.07 \\
Cu & 29 & $-$3.15 &    0.10 & \nodata & \nodata & $-$3.15 & $-$0.83 \\
Zn & 30 & $-$2.16 &    0.08 & \nodata & \nodata & $-$2.16 & $+$0.16 \\
\enddata

\tablenotetext{a}{No solar abundance is available from ionized-species
                  lines.  The [X/Fe] values have been computed using the
                  derived solar abundance from neutral-species lines.}

\end{deluxetable}

\end{center}

\end{document}